\documentclass[12pt]{article}
\usepackage{epsfig}
\usepackage{graphicx}
\usepackage{a4}
\usepackage{latexsym}
\usepackage{cite}

\usepackage{pslatex}
\usepackage[latin1]{inputenc}
\usepackage[T1]{fontenc}

\renewcommand{\theequation}{\thesection.\arabic{equation}}

\newcommand{\hspn}{{\hspace{-4mm}}}
\newcommand{\gsim}{\raisebox{-0.07cm}{$\:\:\stackrel{>}{{\scriptstyle
 \sim}}\:\: $} }

\newcommand{\beq}{\begin{equation}}
\newcommand{\eeq}{\end{equation}}
\newcommand{\bea}{\begin{eqnarray}}
\newcommand{\eea}{\end{eqnarray}}
\newcommand{\nn}{\nonumber}
\newcommand{\ra}{\rightarrow}
\newcommand{\DD}{{\cal D}}
\newcommand{\MSb}{$\overline{\mbox{MS}}$}
\newcommand{\as}{\alpha_{\rm s}}
\newcommand{\ars}{a_{\rm s}}
\newcommand{\ec}{\gamma_e}
\newcommand{\ecs}{\gamma_e^{\,2}}
\newcommand{\ect}{\gamma_e^{\,3}}
\newcommand{\ecf}{\gamma_e^{\,4}}
\newcommand{\ecfi}{\gamma_e^{\,5}}
\newcommand{\ecsi}{\gamma_e^{\,6}}

\newcommand{\Lqr}{L_{\,\rm qr\,}}
\newcommand{\Lqrs}{L_{\,\rm qr\,}^{\,2}}
\newcommand{\Lqrt}{L_{\,\rm qr\,}^{\,3}}
\newcommand{\Lfr}{L_{\,\rm fr\,}}
\newcommand{\Lfrs}{L_{\,\rm fr\,}^{\,2}}
\newcommand{\Lfrt}{L_{\,\rm fr\,}^{\,3}}

\def\z#1{{\zeta_{#1}}}
\def\zs{{\zeta_{2}^{\,2}}}
\def\zt{{\zeta_{2}^{\,3}}}
\def\zzs{{\zeta_{3}^{\,2}}}
\def\ca{{C^{}_A}}
\def\cas{{C^{\,2}_A}}

\def\cf{{C^{}_F}}
\def\cfs{{C^{\, 2}_F}}
\def\cft{{C^{\, 3}_F}}
\def\nf{{n^{}_{\! f}}}
\def\nsq{{n^{\,2}_{\! f}}}
\def\nth{{n^{\,3}_{\! f}}}
\def\dabcnc{{{d^{abc}d_{abc}}\over{n_c}}}

\def\fl11{fl_{11}}

\begin{document}
\setlength{\parskip}{0.2cm}
\setlength{\baselineskip}{0.55cm}

\begin{titlepage}
\noindent
DESY 05-105, SFB/CPP-05-25 \hfill {\tt hep-ph/0506288}\\
DCPT/05/60, $\,\,$IPPP/05/30\\
NIKHEF 05-010 \\[1mm]
June 2005 \\
\vspace{1.5cm}
\begin{center}
\Large
{\bf Higher-Order Corrections in Threshold Resummation} \\
\vspace{2.0cm}
\large
S. Moch$^{\, a}$, J.A.M. Vermaseren$^{\, b}$ and A. Vogt$^{\, c}$\\
\vspace{1.2cm}
\normalsize
{\it $^a$Deutsches Elektronensynchrotron DESY \\
\vspace{0.1cm}
Platanenallee 6, D--15735 Zeuthen, Germany}\\
\vspace{0.5cm}
{\it $^b$NIKHEF Theory Group \\
\vspace{0.1cm}
Kruislaan 409, 1098 SJ Amsterdam, The Netherlands} \\
\vspace{0.5cm}
{\it $^c$IPPP, Department of Physics, University of Durham\\
\vspace{0.1cm}
South Road, Durham DH1 3LE, United Kingdom}\\
\vfill
\large
{\bf Abstract}
\vspace{-0.2cm}
\end{center}
We extend the threshold resummation exponents $G^N$ in Mellin-$N$ space
to the fourth logarithmic (N$^3$LL) order collecting the terms $\as ^{\,
2} (\as \ln N)^n$ to all orders in the strong coupling constant $\as$. 
Comparing the results to our previous three-loop calculations for 
deep-inelastic scattering (DIS), we derive the universal coefficients
$B_{\rm q}$ and $B_{\rm g}$ governing the final-state jet functions 
to order $\as^{\,3}$, extending the previous quark and gluon 
results by one and two orders. A curious relation is found at second
order between these quantities, the splitting functions and the large-%
angle soft emissions in Drell-Yan type processes. We study the numerical
effect of the N$^3$LL corrections using both the fully exponentiated
form and the expansion of the coefficient function in towers of 
logarithms. 
\vspace{1.0cm}
\end{titlepage}
\newpage

%
%
\setcounter{equation}{0}
\section{Introduction}
Coefficient functions, or partonic cross sections, form the backbone of
perturbative QCD. These quantities are defined in terms of power
expansions in the strong coupling constant $\as$. In general, only a 
few terms in this expansion can be calculated. It is however possible, 
and necessary, to resum the dominant contributions to all orders in 
$\as$ close to exceptional kinematic points. Close to threshold, for 
example, where real emissions are kinematically suppressed, the
resummation takes the form of an exponentiation in Mellin-$N$ space 
\cite{Sterman:1987aj,Catani:1989ne,Kidonakis:1997gm,Bonciani:1998vc},
with the moments $N$ defined with respect to the appropriate scaling 
variable, like Bjorken-$x$ in deep-inelastic scattering (DIS) and 
$x_T^{} = 2p_T^{}/ \sqrt{S}$ in direct photon and inclusive hadron 
production.

The resummation exponents are given by integrals over functions in turn
defined by a power series in $\as$. Besides by dedicated calculations, 
the corresponding expansion coefficients can be obtained by expanding 
the exponentials and comparing to the results of fixed-order 
calculations. Hence progress in the latter sector also facilitates 
improved resummation predictions.
At present the next-to-leading order (NLO) is the standard approximation
for many important observables, facilitating a resummation with
next-to-leading logarithmic (NLL) accuracy. For recent introductory 
overviews see, for instance, Refs.~\cite{Laenen:2004pm,Laenen:2005em}. 
The next-to-next-to-leading order (NNLO) corrections have been 
completed so far only for the coefficient functions for inclusive 
lepton-proton DIS
\cite{vanNeerven:1991nn,Zijlstra:1991qc,Zijlstra:1992kj,Moch:1999eb,%
Moch:2004xu}, the Drell-Yan process
\cite{Hamberg:1991np,Harlander:2002wh,Anastasiou:2003ds}
and the related Higgs boson production  
\cite{Harlander:2002wh,Anastasiou:2002yz,Ravindran:2003um,%
Anastasiou:2005qj} in proton-proton collisions.
Consequently, the threshold resummation has been carried out at the
next-to-next-to-leading logarithmic (NNLL) accuracy only for these
processes \cite{Vogt:2000ci,Catani:2003zt}.

Recently we have computed the complete three-loop coefficient functions
for inclusive photon-exchange DIS~\cite{Moch:2004xu,Vermaseren:2005qc}.
Moreover, in the course of the calculation of the third-order splitting
functions governing the NNLO evolution of the parton distributions
\cite{Moch:2004pa,Vogt:2004mw}, we have also computed DIS by exchange
of a scalar directly coupling only to gluons. Together these results
enable us to extend two more universal functions entering the 
resummation exponents, the quark and gluon jet functions $B_{\rm q}$
and $B_{\rm g}$ collecting final-state collinear emissions, to the
third order in $\as$. In fact, already the second-order coefficient for
$B_{\rm g}$ represents a new result, relevant for future NNLL 
resummations of processes with final-state gluons at the Born level.
Making use also of the results of Refs.~\cite{Forte:2002ni,Gardi:2002xm}
we can furthermore effectively, i.e., up to the small contribution of
the four-loop cusp anomalous dimension, extend the threshold 
resummation for inclusive DIS to an unprecedented next-to-next-to-%
next-to-leading logarithmic (N$^3$LL) accuracy.

The remainder of this article is organized as follows: 
after recalling the general structure of the resummation exponents in 
Section~2, we extend the required integrations in Section~3 to the 
fourth logarithmic (N$^3$LL) order. 
In Section~4 we determine the relevant expansion coefficients by 
comparison to our three-loop results for DIS and illustrate the 
numerical effect of the N$^3$LL contributions to the resummation 
exponent. 
In Section~5 we present the resulting predictions for the leading seven
large-$x$ terms of the four-loop coefficient function and discuss 
higher-order effects in terms of the expansion in towers of threshold
logarithms. 
Our results are briefly summarized in Section~6. Some basic relations 
for the integrations of Section 2 can be found in the Appendix.
%
%
\setcounter{equation}{0}
\section{The general structure}
%
%
For processes with only one colour structure at the Born level, the 
resummed Mellin-space coefficient functions $C^{\,N}$ (defined in the 
\MSb\ scheme) are given by a single exponential
\cite{Sterman:1987aj,Catani:1989ne}
\beq
\label{eq:cNres}
  C^{\,N}(Q^2) / C_{\rm LO}^{\,N}(Q^2) \: =\: 
  g_{0}^{}(Q^2) \cdot \exp\, [G^N(Q^2)] \: + \: 
  {\cal O}(N^{-1}\ln^n N) \:\: .
\eeq
Here $C_{\rm LO}^{\,N}$ denotes the lowest-order coefficient function 
for the process under consideration, e.g., $C_{\rm LO}^{\,N} = 1$ for 
DIS.
The prefactor $g_{0}^{}$ collects, order by order in the strong coupling
constant $\as$, all $N$-independent contributions. The exponent $G^N$ 
contains terms of the form $\ln^{\,k} N$ to all orders in $\as$. Besides
the physical hard scale $Q^2$ ($\,= -q^2$ in DIS, with $q$ the four-%
momentum of the exchanged gauge boson), both functions also depend on 
the renormalization scale $\mu_r$ and the mass-factorization scale 
$\mu_{\! f}^{}$. The reference to these scales will be often suppressed 
for brevity.

The exponential in Eq.~(\ref{eq:cNres}) is build up from universal 
radiative factors for each initial- and final-state parton $p$, 
$\Delta_{\,\rm p}$ and $J_{\rm p}$, together with a process-dependent
contribution $\Delta^{\rm int}$. 
For example, the resummation exponents for inclusive deep-inelastic
scattering, Drell-Yan (DY) lepton-pair production and direct photon 
production via $\, q\bar{q} \ra g\gamma\, $ and $\,qg \ra q\gamma\,$ 
\cite{Catani:1998tm} take the form
\bea
\label{eq:GNdec}
  G_{\rm DIS}^N \;\; & = & 
    \ln \Delta_{\,\rm q} \: + \: \ln J_{\rm q} \: + \: 
    \ln \Delta^{\,\rm int}_{\,\rm DIS} \:\: , \nn \\[1mm]
  G_{\rm DY}^N \;\;\;  & = & 
    2\, \ln \Delta_{\,\rm q} \: + \: 
    \ln \Delta^{\,\rm int}_{\,\rm DY} \:\: , \nn \\[1mm]
  G_{\rm ab\ra c\gamma}^N\!\!  & = &
    \ln \Delta_{\,\rm a} \: + \: \ln \Delta_{\,\rm b} \: + \:
    \ln J_{\rm c} \: + \: \ln \Delta^{\,\rm int}_{\,\rm ab\ra c\gamma} 
    \:\: .
\eea
The radiation factors are given by integrals over functions of the 
running coupling. Specifically, the effects of collinear soft-gluon 
radiation off an initial-state parton $p = q,g$ are collected by
\beq
\label{eq:dint}
  \ln \Delta_{\,\rm p} (Q^2,\, \mu_f^{\,2}) \: = \: \int_0^1 \! dz \,
  \frac{z^{N-1}-1}{1-z} \,\int_{\mu_f^{\,2}}^{(1-z)^2 Q^2}
  \frac{dq^2}{q^2}\, A_{\rm p}(\as(q^2)) \:\: .
\eeq
Collinear emissions from an `unobserved' final-state parton lead to
the so-called jet function,
\beq
\label{eq:Jint}
  \ln J_{\rm p} (Q^2) \: = \: \int_0^1 \! dz \,\frac{z^{N-1}-1}{1-z}
  \, \left[ \int_{(1-z)^2 Q^2}^{(1-z) Q^2} \frac{dq^2}{q^2}\,
  A_{\rm p}(\as(q^2)) + B_{\rm p} (\as([1-z] Q^2)) \right] \:\: .
\eeq
Finally the process-dependent contributions from large-angle soft 
gluons are resummed by
\beq
\label{eq:Dint}
  \ln \Delta^{\rm int} (Q^2) \: = \: \int_0^1 \! dz
  \,\frac{z^{N-1}-1}{1-z} \, D (\as([1-z]^2 Q^2)) \:\: .
\eeq
The functions $g_{0}^{}$ in Eq.~(\ref{eq:cNres}) and 
$A_{\rm p}$, $B_{\rm p}$ and $D$ in Eqs.~(\ref{eq:dint}) -- (\ref
{eq:Dint}) are given by the expansions
\beq
\label{eq:aexp}
  F(\as) \:\: = \:\: 
  \sum_{l=l_0}^{\infty}\, F_l\: \frac{\as^{\,l}}{4\pi} \:\: \equiv \:\: 
  \sum_{l=l_0}^{\infty}\, F_l\: \ars^{\,l} \:\: ,
\eeq
where $l_0= 0$ with $g_{00}^{}= 1$ for $F= g_0^{}$, and $l_0= 1$ else.
Here we have also taken the opportunity to specify the reduced coupling
$a_{\rm s}$ employed for the rest of this article. The extent to which 
these functions are known defines the accuracy to which the threshold 
logarithms can be resummed.

The situation for inclusive DIS is actually simpler than indicated in
Eq.~(\ref{eq:GNdec}), as the function (\ref {eq:Dint}) is found to 
vanish to all orders in $\as$ \cite{Forte:2002ni,Gardi:2002xm},
\beq
\label{eq:DDIS}
   D^{\,\rm DIS}_k \; = \; 0 \:\: , \quad
   \Delta_{\,\rm DIS} ^{\,\rm int} \; = \; 1 \:\: . 
\eeq
On the other hand, the resummation of processes with four or more 
partons at the Born level, like inclusive hadron production in $pp$ 
collisions \cite{deFlorian:2005yj}, is more complicated than Eq.~(\ref
{eq:cNres}) due to colour interferences and correlations in the large-%
angle soft gluon emissions \cite{Kidonakis:1997gm,Bonciani:1998vc}. 
However, the process-independent functions $A_{\rm p}$ and $B_{\rm p}$ 
retain their relevance also for such cases. 
%
%
\setcounter{equation}{0}
\section{The resummation exponent to fourth logarithmic order}
%
%
After performing the integrations in Eqs.~(\ref{eq:dint}) -- 
(\ref{eq:Dint}), the function $G^N$ in Eq.~(\ref{eq:GNdec}) takes the 
form
\beq
\label{eq:GNexp}
  G^N(Q^2) \: = \: 
  \ln N \cdot g_1^{}(\lambda) \: + \: g_2^{}(\lambda) \: + \: 
  \ars\, g_3^{}(\lambda) \: + \: \ars^{\,2}\, g_4^{}(\lambda) \: + \: 
  \ldots \:\: ,
\eeq
%
where $\lambda = \beta_0\, \ars\, \ln N$.
For the actual computation of the functions $g_i^{}$ it is convenient 
to employ the following representation for the scale dependence of 
$\ars$ up to N$^3$LO$\,$:
\bea
\label{eq:arun}
\ars(q^2) &=& 
    {\ars \over X} 
  \: - \: {\ars^{\,2} \over X^2} {\beta_1 \over \beta_0} \ln X 
  \: + \: {\ars^{\,3} \over X^3} \left[ 
    {\beta_1^2 \over \beta_0^2} (\ln^2 X - \ln X - 1 + X) 
    + {\beta_2 \over \beta_0} (1-X) 
    \right]
\nonumber \\
& & \mbox{}
  \: + \: {\ars^{\,4} \over X^4} \left[ 
    {\beta_1^3 \over \beta_0^3} \left(
    2 (1-X) \ln X + {5 \over 2} \ln^2 X - \ln^3 X - {1 \over 2} 
    + X - {1 \over 2} X^2 \right)
    \right.
\nonumber \\
& & \left. \mbox{}
    + {\beta_3 \over 2 \beta_0} (1-X^2) 
    + {\beta_1 \beta_2 \over \beta_0^2} (2 X \ln X - 3 \ln X - X (1-X)) 
    \right]
  \: + \: {\cal O} \big( \ars^{\,5} \big)
\eea
with $\ars \equiv \ars(\mu_r^{\,2})$ and the abbreviation
$  X = 1 + \ars \beta_0 \ln (q^2 / \mu_r^{\,2}) \, .  $
The terms up to the $n$-th order in $\ars$ in Eq.~(\ref{eq:arun}) 
contribute to $g_n^{}$ in Eq.~(\ref{eq:GNexp}). Thus the calculation of
$g_4^{}$ requires the highest known coefficient of the beta function of 
QCD, $\beta_3$~\cite{vanRitbergen:1997va,Czakon:2004bu}.

Generalizing the approach of Ref.~\cite{Vogt:2000ci},
the functions $g_i^{}(\lambda)$ can be obtained using well-known methods
for Mellin transforms based on properties of harmonic sums and harmonic 
polylogarithms~\cite{Vermaseren:1998uu,Remiddi:1999ew} in addition to
algorithms for the evaluation of nested sums~\cite{Moch:2001zr}. The 
basic relations for this approach, suitable for the evaluation of 
Eq.~(\ref{eq:GNexp}) to any accuracy, are presented in the Appendix. As 
a check we have also carried out the integrations along the lines of 
Ref.~\cite{Catani:2003zt}.

For the convenience of the reader, we first recall the known results for
$g_{1}^{}$, $g_{2}^{}$ and $g_{3}^{}$ 
\cite{Sterman:1987aj,Catani:1989ne,Vogt:2000ci,Catani:2003zt}.
For brevity suppressing factors of $\beta_0$ (see below) and using the 
short-hand notations $\,L_{\rm qr} = \ln (Q^2/\mu_r^{\,2})\,$ and  
$\,L_{\rm fr} = \ln (\mu_f^{\,2}/\mu_r^{\,2})\,$, these functions can 
be written as
\bea
  g_1^{\,\rm DIS}(\lambda) & = & 
          A_1 \*  (
            1
          - \ln(1-\lambda)
          + \lambda^{-1} \* \ln(1-\lambda) 
          )
\label{eq:g1n}
\:\: ,
\\
  g_2^{\,\rm DIS}(\lambda) & = & 
        \bigl(
            A_1 \* \beta_1 
          - A_2
        \bigr) \* (
            \lambda
          + \ln(1-\lambda)
          )
          + {1 \over 2} \* A_1 \* \beta_1 \* \ln^2(1-\lambda)
\nonumber\\
&&\mbox{}
       - \bigl( A_1 \* \ec - B_1 \bigr) \* \ln(1-\lambda)
       + \Lqr \* A_1 \* \ln(1-\lambda)
       + \Lfr \* A_1 \* \lambda
\label{eq:g2n}
\:\: ,
\\[1ex]
  g_3^{\,\rm DIS}(\lambda) & = & 
         {1 \over 2} \* (
            A_1 \* \beta_2 
          - A_1 \* \beta_1^2 
          + A_2 \* \beta_1 
          - A_3 
         ) \*  \biggl(
            1
          + \lambda
          - {1 \over 1-\lambda}
          \biggr)
\nonumber\\
&&\mbox{}
      +  A_1 \* \beta_1^2 \*  \biggl(
            {\ln(1-\lambda) \over 1-\lambda}
          + {1 \over 2} \* {\ln^2(1-\lambda) \over 1-\lambda}
          \biggr)
      + \biggl(
            A_1 \* \beta_2 
          - A_1 \* \beta_1^2 
         \biggr) \* \ln(1-\lambda)
\nonumber\\
&&\mbox{}
       +  (
            A_1 \* \beta_1 \* \ec 
          + A_2 \* \beta_1 
          - B_1 \* \beta_1 
          ) \* \biggl(
            1
          - {1 \over 1-\lambda} 
          - {\ln(1-\lambda) \over 1-\lambda}
          \biggr)
\nonumber\\
&&\mbox{}
       - \biggl(
            A_1 \* \beta_2 
          + {1 \over 2} \* A_1  \* (\ecs + \z2) 
          + A_2 \* \ec
          - B_1 \* \ec
          - B_2 
         \biggr) \* \biggl(
            1
          - {1 \over 1-\lambda}
          \biggr)
\nonumber\\
&&\mbox{}
       + \Lqr  \*  \biggl[
         (
            A_1 \* \ec 
          - A_1 \* \beta_1 
          + A_2 
          - B_1 
          ) \*  \biggl(
            1 
          - {1 \over 1-\lambda}
          \biggr)
       +  A_1 \* \beta_1 \* \biggl(
          {\ln(1-\lambda) \over 1-\lambda}
          \biggr)
          \biggl]
\nonumber\\
&&\mbox{}
       + \Lfr \* A_2 \* \lambda 
       - \Lqrs \* {1 \over 2} \* A_1 \*  \biggl(
            1
          - {1 \over 1-\lambda}
          \biggr)
       - \Lfrs \* {1 \over 2} \* A_1 \* \lambda 
\label{eq:g3n}
\:\: .
\eea
The dependence on $\beta_0$ is recovered by $A_k \ra A_k /
\beta_0^{\,k}$, $B_k \ra B_k /\beta_0^{\, k}$, $\beta_k \ra \beta_k 
/\beta_0^{\,k+1}$ and multiplication of $g_3$ by $\beta_0$.
In the same notation the new function $g_4$ (to be multiplied by
$\beta_0^{\,2}$) is given by 
\bea
  g_4^{\rm DIS}(\lambda) & \! = \! & 
       - {1 \over 6} \* A_1 \* \beta_1^3 \* 
         {\ln^3(1-\lambda) \over (1-\lambda)^2}
       + {1 \over 2} \* (
         A_1 \* \beta_1^2 \* \ec
       + A_2 \* \beta_1^2 
       - B_1 \* \beta_1^2 
       ) \* {\ln^2(1-\lambda) \over (1-\lambda)^2}
       + {1 \over 2} \* (
         A_1 \* \beta_1^3  
       - A_1 \* \beta_1 \* \beta_2 
\nonumber\\
&&\mbox{}
       - A_1 \* \beta_1 \* (\ecs + \z2) 
       + A_2 \* \beta_1^2 
       - 2 \* A_2 \* \beta_1 \* \ec
       - A_3 \* \beta_1 
       + 2 \* B_1 \* \beta_1 \* \ec 
       + 2 \* B_2 \* \beta_1 
       ) \* {\ln(1-\lambda) \over (1-\lambda)^2}
\nonumber\\
&&\mbox{}
       - (
         A_1 \* \beta_1^3 
       - A_1 \* \beta_1 \* \beta_2  
       ) \* {\ln(1-\lambda) \over 1-\lambda}
       + \biggl(
         {1 \over 2} \* A_1 \* \beta_1^3 
       - A_1 \* \beta_1 \* \beta_2  
       + {1 \over 2} \* A_1 \* \beta_3
       \biggr) \* \ln(1-\lambda)
       + (
         A_1 \* \beta_1^3 
\nonumber\\
&&\mbox{}
       - A_1 \* \beta_1 \* \beta_2  
       - A_1 \* \beta_1^2 \* \ec
       + A_1 \* \beta_2  \* \ec
       - A_2 \* \beta_1^2 
       + A_2 \* \beta_2 
       + B_1 \* \beta_1^2 
       - B_1 \* \beta_2 
       ) \* \biggl(
            {1 \over 2} 
          - {1 \over 1-\lambda} 
\nonumber\\
&&\mbox{}
          + {1 \over 2} \* {1 \over (1-\lambda)^2} 
          \biggr)
       + {1 \over 2} \* \biggl(
         {1 \over 3} \* A_1 \* \beta_1^3 
       - {1 \over 6} \* A_1 \* \beta_1 \* \beta_2 
       - {1 \over 6} \* A_1 \* \beta_3 
       - {1 \over 3} \* A_1 \* (3\* \ec \* \z2 
                 + \ect + 2 \* \z3)
\nonumber\\
&&\mbox{}
       + A_2 \* \beta_1 \* \ec
       - A_2 \* (\ecs + \z2)
       - {5 \over 6} \* A_2 \* \beta_1^2 
       + {1 \over 3} \* A_2 \* \beta_2 
       + {5 \over 6} \* A_3 \* \beta_1 
       - A_3 \* \ec
       - {1 \over 3} \* A_4 
       - B_2 \* \beta_1 
\nonumber\\
&&\mbox{}
       + B_1 \* (\ecs + \z2)
       + 2 \* B_2 \* \ec 
       + B_3 
       \biggr) \* \biggl(
            1
          - {1 \over (1-\lambda)^2}
          \biggr)
       + {1 \over 3} \* \biggl(
         A_1 \* \beta_1^3  
       - 2 \* A_1 \* \beta_1 \* \beta_2  
       + A_1 \* \beta_3 
\nonumber\\
&&\mbox{}
       + A_2 \* \beta_2 
       - A_2 \* \beta_1^2 
       + A_3 \* \beta_1 
       - A_4 \biggr) \* \lambda
       + \Lqr  \*  \biggl[
         (
         A_1 \* \beta_1^2 
       - A_1 \* \beta_2 
       ) \* \biggl(
            {1 \over 2} 
          - {1 \over 1-\lambda} 
          + {1 \over 2} \* {1 \over (1-\lambda)^2} 
          \biggr)
\nonumber\\
&&\mbox{}
       + \biggl(
         {1 \over 2} \* A_1 \* (\ecs + \z2)
       - {1 \over 2} \* A_2 \* \beta_1 
       + A_2 \* \ec
       + {1 \over 2} \* A_3 
       - B_1 \* \ec 
       - B_2 
       \biggr) \* \biggl(
            1
          - {1 \over (1-\lambda)^2}
          \biggr)
\nonumber\\
&&\mbox{}
       + (
         A_1 \* \beta_1 \* \ec 
       + A_2 \* \beta_1 
       - B_1 \* \beta_1  
       ) \* {\ln(1-\lambda) \over (1-\lambda)^2}
       - {1 \over 2} \* A_1 \* \beta_1^2 \* 
         {\ln^2(1-\lambda) \over (1-\lambda)^2}
          \biggr]
       - \Lqrs \*  \biggl[
         {1 \over 2} \* (
            A_1 \* \ec 
\nonumber\\
&&\mbox{}
          + A_2 
          - B_1 
          ) \*  \biggl(
            1
          - {1 \over (1-\lambda)^2}
          \biggr)
          + {1 \over 2} \* A_1 \* \beta_1 \* 
            {\ln(1-\lambda) \over (1-\lambda)^2}
          \biggr]
       +  \Lqrt \* {1 \over 6} \* A_1 \* \biggl(
            1
          - {1 \over (1-\lambda)^2}
          \biggr)
\nonumber\\
&&\mbox{}
       + \Lfr \* A_3 \* \lambda
       - \Lfrs \*  \biggl(
            A_2
       +   {1 \over 2} \* A_1 \* \beta_1 
       \biggl) \* \lambda
       + \Lfrt \* {1 \over 3} \* A_1 \* \lambda
\label{eq:g4n}
\, .
\eea
The results $g_{i}^{\,\rm DY}$ for the Drell-Yan process and, with 
slightly different coefficients $A_i$ and $D_i$, Higgs production 
via gluon-gluon fusion, are related to 
Eqs.~(\ref{eq:g1n}) -- (\ref{eq:g4n}) as follows: the function 
corresponding to Eq.~(\ref{eq:g1n}) reads 
$\, g_1^{\,\rm DY}(\lambda) = 2\, g_1^{\,\rm DIS} (2\lambda)\,$, 
while the functions $g_2^{\,\rm DY}, g_3^{\,\rm DY}$ and 
$g_4^{\,\rm DY}$ are obtained from Eqs.~(\ref{eq:g2n}) -- (\ref{eq:g4n})
by replacing $\lambda \rightarrow 2\lambda$ everywhere and substituting
$B_i \rightarrow D_{i}/2$ in all terms. Finally, the constants have to 
be changed according to $\ec \rightarrow 2\ec$ and 
$\zeta_n \rightarrow 2^n \zeta_n$. 
The generalization of Eqs.~(\ref{eq:g1n}) -- (\ref{eq:g4n}) to other 
processes involving Eqs.~(\ref{eq:dint}) -- (\ref{eq:Dint}) is obvious. 

The functions $g_1^{}(\lambda)$ collecting the leading logarithms $\,L 
(\ars L)^k\,$ depend on $A_1$ only and are finite for all $\lambda$. The
N$^{\,n-1}$LL contributions $g_{n>1}(\lambda)$ to Eq.~(\ref{eq:GNexp})
including $A_n$, $B_{n-1}$ and $D_{n-1}$, on the other hand, exhibit
Landau poles at the moments $\,N = \exp [1/(\beta_0 a_s)]\,$ for DIS 
and $\,N = \exp [1/(2\beta_0 a_s)]\,$ for the DY case (and at both 
values in general). 
 As obvious from Eq.~(\ref{eq:g1n}) -- (\ref{eq:g4n}) the strength of 
these singularities increases with the logarithmic order, reaching
$[1-(2)\lambda]^{-2}$ at the N$^3$LL level. 
%
%
\setcounter{equation}{0}
\vspace*{-1mm}
\section{Resummation coefficients and numerical stability}
%
%
Since observables are independent, order by order in $\as$, of the 
factorization scale, the functions $A_{\rm p}$ in Eqs.~(\ref{eq:dint})
and (\ref{eq:Jint}) are given by the large-$N$ coefficients of the 
diagonal splitting functions for $\mu_r = \mu_{\! f}^{}$, 
\beq
  P_{\rm pp}(\as) \; = \; - A_{\rm p}(\as)\: \ln N + {\cal O}(1)
  \:\: ,
\eeq
which in turn are identical to the anomalous dimension of a Wilson line
with a cusp~\cite{Korchemsky:1989si}. The first and second order 
coefficients have been known for a long time, the third order has been
recently completed by us. The expansion coefficients (\ref{eq:aexp}) 
for the quark case read~\cite{Kodaira:1981nh,Moch:2004pa}
\bea
\label{eq:Aqexp}
  A_{\rm q,1} & \! = \! & 4\, C_F \nn \\[0.5mm]
  A_{\rm q,2} & \! = \! & 8\, C_F \left[ \left( \frac{67}{18} 
     - \zeta_2^{} \right) C_A - \frac{5}{9}\,\nf \right] \nn \\[1mm]
  A_{\rm q,3} & \! = \! &
     16\, C_F \left[ C_A^{\,2} \,\left( \frac{245}{24} - \frac{67}{9}
     \: \z2 + \frac{11}{6}\:\z3 + \frac{11}{5}\:\zs \right) 
   \: + \: C_F \nf\, \left( -  \frac{55}{24}  + 2\:\z3
   \right) \right. \nn\\ & & \left. \mbox{} \qquad
   + \: C_A \nf\, \left( - \frac{209}{108}
         + \frac{10}{9}\:\z2 - \frac{7}{3}\:\z3 \right)
   \: + \: \nsq \left( - \frac{1}{27}\,\right) \right] \:\: .
\eea
Here $\nf$ denotes the number of effectively massless quark flavours, 
and $C_F$ and $C_A$ are the usual colour factors, with $C_F = 4/3$ and 
$C_A = 3$ in QCD.  The gluonic quantities are given by
\beq
\label{eq:Ag}
 A_{\rm g,i} \; = \; C_A / C_F \; A_{\rm q,i} \:\: .
\eeq
The perturbative expansion of $A_{\rm p}$ is very benign. For $\nf = 
4$, for example, Eqs.~(\ref{eq:Aqexp}) lead to
\beq
\label{Aqnum}
  A_{\rm q}(\as) \; \cong \; 0.4244\:\as \: 
  ( 1 \: + \: 0.6381\: \as \: + \: 0.5100\: \as^2 \: + \: \ldots ) 
  \:\: . 
\eeq
Consequently, already the effect of $A_3$ on the resummed coefficient
functions is very small~\cite{Vogt:2000ci,Catani:2003zt}, and a simple 
estimate suffices for the presently unknown fourth-order coefficients 
$A_4$ entering $g_4^{}$. With the [0/2] results differing by less than
10\%, we will employ the [1/1] Pad\'e approximants 
\bea
\label{eq:A4pade}
  A_{\rm q,4} \; \approx \; 7849\: ,\:\: 4313\: ,\:\:  1553 \quad
  \mbox{for} \quad \nf \; = \; 3\: ,\:\: 4\: ,\:\: 5 \:\: ,
\eea
corresponding to an estimate of $\, +\:0.4075\:\as^{3}\, $ for the next
term in Eq.~(\ref{Aqnum}).

\pagebreak

For the determination of the coefficients $B_i$ we also need the
constant-$N$ piece, $g_0^{}$ in Eq.~(\ref{eq:cNres}), for inclusive 
DIS. The corresponding expansion coefficients $g_{0k}^{}$ can be 
obtained by Mellin inverting the +-distribution and $\delta(1-x)$ parts 
of the $k$-th order coefficient function $c^{\,(k)}_{\,2,\rm q}$. Again
using the expansion parameter $\ars = \as /(4\pi)$, the presently known 
terms \cite{vanNeerven:1991nn,Moch:1999eb,Vermaseren:2005qc} are 
given by 
\bea
\label{eq:g01dis}
  g_{01}^{\rm DIS} & = & \cf \* \left( - 9 - 2 \* \z2 
  + 2 \* \ecs + 3 \* \ec \right)
 \:\: ,
\\[0.5mm]
\label{eq:g02dis}
  g_{02}^{\rm DIS} &=& 
         \cfs  \*  \left( {331 \over 8} - {51 \over 2} \* \ec 
       - {27 \over 2} \* \ecs + 6 \* \ect + 2 \* \ecf 
       + {111 \over 2} \* \z2 - 18 \* \ec \* \z2 - 4 \* \ecs \* \z2 
\right.
\nonumber\\
&&\mbox{}
\left.
       - 66 \* \z3 + 24 \* \ec \* \z3 
       + {4 \over 5} \* \zs \right)
     + \ca \* \cf  \*  \left(  - {5465 \over 72} 
       + {3155 \over 54} \* \ec + {367 \over 18} \* \ecs 
\right.
\nonumber\\
&&\mbox{}
\left.
       + {22 \over 9} \* \ect 
       - {1139 \over 18} \* \z2 - {22 \over 3} \* \ec \* \z2 
       - 4 \* \ecs \* \z2 
       + {464 \over 9} \* \z3 - 40 \* \ec \* \z3 
       + {51 \over 5} \* \zs  \right)
\nonumber\\
&&\mbox{}
     + \cf \* \nf  \*  \left( {457 \over 36} - {247 \over 27} \* \ec 
       - {29 \over 9} \* \ecs - {4 \over 9} \* \ect 
       + {85 \over 9} \* \z2 + {4 \over 3} \* \ec \* \z2 
       + {4 \over 9} \* \z3 \right)
\eea
and
\bea
\label{eq:g03dis}
  g_{03}^{\rm DIS} &=& 
       \cft  \*  \left(  - {7255 \over 24} 
         + {1001 \over 8} \* \ec + {187 \over 4} \* \ecs 
         - {93 \over 2} \* \ect - 9 \* \ecf + 6 \* \ecfi 
         + {4 \over 3} \* \ecsi - {6197 \over 12} \* \z2 
\right.
\nonumber\\
&&\mbox{}
         + {579 \over 2} \* \ec \* \z2 + 66 \* \ecs \* \z2 
         - 36 \* \ect \* \z2 - 4 \* \ecf \* \z2 
         - 411 \* \z3 - 346 \* \ec \* \z3 - 60 \* \ecs \* \z3 
\nonumber\\
&&\mbox{}
         + 48 \* \ect \* \z3 - {1791 \over 5} \* \zs 
         + 84 \* \ec \* \zs + {8 \over 5} \* \ecs \* \zs 
         + 556 \* \z2 \* \z3 - 80 \* \ec \* \z2 \* \z3 
         + 1384 \* \z5 
\nonumber\\
&&\mbox{}
\left. 
         - 240 \* \ec \* \z5 
         + {8144 \over 315} \* \zt - {176 \over 3} \* \zzs \right)
       + \ca \* \cfs  \*  \left( {9161 \over 12} 
         - {16981 \over 24} \* \ec - {5563 \over 36} \* \ecs
\right. 
\nonumber\\
&&\mbox{} 
         + {8425 \over 54} \* \ect + {433 \over 9} \* \ecf 
         + {44 \over 9} \* \ecfi 
         + {191545 \over 108} \* \z2 - {28495 \over 54} \* \ec \* \z2 
         - {592 \over 3} \* \ecs \* \z2 - 8 \* \ecf \* \z2 
\nonumber\\
&&\mbox{}
         - {284 \over 9} \* \ect \* \z2 
         - {49346 \over 27} \* \z3 + 752 \* \ec \* \z3 
         + {640 \over 9} \* \ecs \* \z3 - 80 \* \ect \* \z3 
         + {11419 \over 27} \* \zs 
\nonumber\\
&&\mbox{}
         + {299 \over 3} \* \ec \* \zs 
         + {142 \over 5} \* \ecs \* \zs - 828 \* \z2 \* \z3 
         + 96 \* \ec \* \z2 \* \z3 - {3896 \over 9} \* \z5 
         + 120 \* \ec \* \z5 
\nonumber\\
&&\mbox{}
\left.
         - {23098 \over 315} \* \zt + {536 \over 3} \* \zzs \right)
       + \cas \* \cf  \*  \left(  - {1909753 \over 1944} 
         + {599375 \over 729} \* \ec + {50689 \over 162} \* \ecs 
\right. 
\nonumber\\
&&\mbox{}
         + {4649 \over 81} \* \ect + {121 \over 27} \* \ecf 
         - {78607 \over 54} \* \z2 - {18179 \over 81} \* \ec \* \z2 
         - {778 \over 9} \* \ecs \* \z2 - {88 \over 9} \* \ect \* \z2 
\nonumber\\
&&\mbox{}
         + {115010 \over 81} \* \z3
         + {121 \over 27} \* \ecf 
         - {78607 \over 54} \* \z2 - {18179 \over 81} \* \ec \* \z2 
         - {778 \over 9} \* \ecs \* \z2 - {88 \over 9} \* \ect \* \z2 
\nonumber\\
&&\mbox{}
\left. 
         + {3496 \over 9} \* \z2 \* \z3 
         + {176 \over 3} \* \ec \* \z2 \* \z3 
         - {416 \over 3} \* \z5 + 232 \* \ec \* \z5 
         - {12016 \over 315} \* \zt - {248 \over 3} \* \zzs \right)
\nonumber\\
&&\mbox{}
       + \cfs \* \nf  \*  \left(  - {341 \over 36} 
         + {2003 \over 108} \* \ec + {83 \over 18} \* \ecs 
         - {683 \over 27} \* \ect - {70 \over 9} \* \ecf 
         - {8 \over 9} \* \ecfi - {10733 \over 54} \* \z2 
\right.
\nonumber\\
&&\mbox{}
         + {2177 \over 27} \* \ec \* \z2 
         + {112 \over 3} \* \ecs \* \z2 + {32 \over 9} \* \ect \* \z2
         + {10766 \over 27} \* \z3 - {20 \over 9} \* \ec \* \z3 
         + {8 \over 9} \* \ecs \* \z3 
         - {8 \over 3} \* \ec \* \zs 
\nonumber\\
&&\mbox{}
\left.
         - {10802 \over 135} \* \zs 
         - {40 \over 3} \* \z2 \* \z3 - {784 \over 9} \* \z5 \right)
       + \ca \* \cf \* \nf  \*  \left( {142883 \over 486} 
         - {160906 \over 729} \* \ec 
\right.
\nonumber\\
&&\mbox{}
         - {7531 \over 81} \* \ecs 
         - {1552 \over 81} \* \ect - {44 \over 27} \* \ecf 
         + {33331 \over 81} \* \z2 + {5264 \over 81} \* \ec \* \z2 
         + {56 \over 3} \* \ecs \* \z2 + {16 \over 9} \* \ect \* \z2 
\nonumber\\
&&\mbox{}
\left.
         - {21418 \over 81} \* \z3 + {1976 \over 27} \* \ec \* \z3 
         + 8 \* \ecs \* \z3 + {164 \over 135} \* \zs 
         - {128 \over 15} \* \ec \* \zs 
         - {64 \over 9} \* \z2 \* \z3 + {8 \over 3} \* \z5 \right)
\nonumber\\
&&\mbox{}
       + \cf \* \nsq  \*  \left(  
         - {9517 \over 486} 
         + {8714 \over 729} \* \ec + {470 \over 81} \* \ecs 
         + {116 \over 81} \* \ect + {4 \over 27} \* \ecf
         - {2110 \over 81} \* \z2 - {8 \over 9} \* \ecs \* \z2 
\right.
\nonumber\\
&&\mbox{}
\left.
         - {116 \over 27} \* \ec \* \z2 
         + {80 \over 81} \* \z3 + {64 \over 27} \* \ec \* \z3 
         - {292 \over 135} \* \zs \right)
       + \dabcnc \* fl_{11} \* \left( 64 + 160 \* \z2 
\right.
\nonumber\\
&&\mbox{}
\left.
         + {224 \over 3} \* \z3 
         - {32 \over 5} \* \zs - {1280 \over 3} \* \z5 \right)
\, .
\eea
Note the new flavour structure $fl_{11}$
\cite{Vermaseren:2005qc,Larin:1997wd} in $g_{03}^{}$. This contribution, 
introducing the colour factor $d^{abc}d_{abc}/n_c$, for the first time 
leads to a difference between the flavour-singlet and non-singlet 
coefficient functions for the photon-exchange structure function $F_2$ 
in the soft-gluon limit, with $fl_{11}^{\,\rm ns} = 3\langle e\rangle\,$
and $fl_{11}^{\,\rm s} = \langle e \rangle ^2 /\langle e^{\,2} \rangle\,
$, where $\langle e^{\,k} \rangle$ represents the average of the charge
 $e^{\,k}$ for the active quark flavours, $\langle e^{\,k} \rangle = 
n_{\! f}^{-1} \sum_{\,i=1}^{\,n_{\! f}}\, e_i^{\,k}$. 
Correspondingly, the large-$N$ coefficient functions for $Z$- and 
$W$-exchange DIS will differ from each other and from $F_2^{\,\rm e.m.}$
at order $\as^3$. Based on the size of the $fl_{11}$ term in Eq.~(\ref
{eq:g03dis}), however, we expect these differences to be numerically 
insignificant.

Now the coefficients $B_{{\rm q},k}$ entering the jet function (\ref
{eq:Jint}) can be derived successively from the $\ln\, N\,$ terms of 
the $k$-loop DIS coefficient functions $c_{\,2,\rm q}^{\,(k)}$. 
Expansion of Eqs.~(\ref{eq:GNexp}) and (\ref{eq:g1n}) -- (\ref{eq:g4n}) 
in powers of $\ars$ yields
\bea
  c_{\,2,\rm q}^{\,(1)} \Big|_{\,\ln N} & = & A_1 \,\ec - B_1 \nn\\[1mm]
  c_{\,2,\rm q}^{\,(2)} \Big|_{\,\ln N} & = &
   \frac{1}{2} A_1 \beta_0 \, ( \ecs + \z2 ) + A_2\,\ec - B_1 \beta_0\ec 
   - B_2 + g_{01}^{} ( A_1 \ec - B_1 )  \nn \\[1mm]
  c_{\,2,\rm q}^{\,(3)} \Big|_{\,\ln N} & = &
    \frac{1}{3} A_1 \beta_0^2 ( \ect + 3 \ec \z2 + 2 \z3 )
  + \frac{1}{2} A_1 \beta_1 ( \ecs + \z2 )
  + A_2 \beta_0 ( \ecs + \z2 ) \nn \\ & & \mbox{}
  + A_3 \ec - B_1 ( \beta_1 \ec + \beta_0^2 \z2 + \beta_0^2 \ecs) 
  - 2 B_2 \beta_0 \ec - B_3 \nn \\ & & \mbox{}
  + g_{02}^{} ( A_1 \ec - B_1 ) + g_{01}^{} \bigg( \frac{1}{2} A_1 \
     \beta_0 ( \ecs + \z2 ) + A_2\,\ec - B_1 \beta_0 \ec - B_2 \bigg) 
  \:\: .
\label{eq:cklnN}
\eea
The coefficients of $\,\ln^{\,l} N\,$, $2\leq\,l \leq 2k\,$ in $c_{\,2,
\rm q}^{\,(k)}$, on the other hand, are completely fixed by lower-order 
resummation coefficients, thus providing an explicit $k$-loop check of 
the exponentiation formula. Comparison of the relations (\ref{eq:cklnN})
with the corresponding results from the fixed-order calculations of 
Refs.~\cite{vanNeerven:1991nn,Moch:1999eb,Vermaseren:2005qc}, using
Eqs.~(\ref{eq:Aqexp}), (\ref{eq:g01dis}) and (\ref{eq:g02dis}), 
leads to
\bea
\label{eq:B1}
  B_{\rm q,1} & \! =\!& - 3\: \cf 
\:\: ,
\\[1mm]
\label{eq:B2}
  B_{\rm q,2}  &\! =\!&
  \cfs \* \left[ - {3 \over 2} + 12\: \* \z2 - 24\: \* \z3 \right]
  + \cf \* \ca \* \left[ - {3155 \over 54} + {44 \over 3}\: \* \z2 
  + 40\: \* \z3 \right] 
  \nn \\ & & \mbox{\hspn}
  + \cf \* \nf \* \left[ {247 \over 27} - {8 \over 3}\: \* \z2 \right]
\:\: ,
\\[1mm]
\label{eq:B3}
  B_{\rm q,3}  &\! =\! & 
       \cft  \*  \left[  - {29 \over 2}\: - 18\: \* \z2 - 68\: \* \z3 
         - {288 \over 5}\: \* \zs + 32\: \* \z2 \* \z3 
         + 240\: \* \z5 \right]
\nonumber\\[0.5mm]
&&\mbox{\hspn}
       + \ca \* \cfs  \*  \left[  - 46 + 287\: \* \z2 
         - {712 \over 3}\: \* \z3 - {272 \over 5}\: \* \zs 
         - 16\: \* \z2 \* \z3 - 120\: \* \z5 \right]
\nonumber\\[0.5mm]
&&\mbox{\hspn}
       + \cas \* \cf  \*  \left[  - {599375 \over 729} 
         + {32126 \over 81}\: \* \z2 + {21032 \over 27}\: \* \z3 
         - {652 \over 15}\: \* \zs - {176 \over 3}\: \* \z2 \* \z3 
         - 232\: \* \z5 \right]
\nonumber\\[0.5mm]
&&\mbox{\hspn}
       + \cfs \* \nf  \*  \left[ {5501 \over 54} - 50\: \* \z2 
         + {32 \over 9}\: \* \z3 \right]
       + \cf \* \nsq  \*  \left[  - {8714 \over 729} 
         + {232 \over 27}\: \* \z2 - {32 \over 27}\: \* \z3 \right]
\nonumber\\[0.5mm]
&&\mbox{\hspn}
       + \ca \* \cf \* \nf  \*  \left[ {160906 \over 729} 
         - {9920 \over 81}\: \* \z2 - {776 \over 9}\: \* \z3 
         + {208 \over 15}\: \* \zs \right]
\:\: .
\eea
Eq.~(\ref{eq:B1}) is, of course, a well-known result 
\cite{Sterman:1987aj,Catani:1989ne}. Eq.~(\ref{eq:B2}) has been derived
by us before \cite{Moch:2002sn}, establishing $D_2^{\,\rm DIS} = 0$ 
from the $\nf \ln^2 N$ term at three loops. For our new result 
(\ref{eq:B3}), on the other hand, we have to rely on the subsequent 
all-order proofs of Eq.~(\ref{eq:DDIS}) in Refs.~\cite{Forte:2002ni,%
Gardi:2002xm}. The QCD expansion of $B_{\rm q}$ analogous to 
Eq.~(\ref{Aqnum}) appears far less stable than that for $A_{\rm q}$, 
\beq
\label{eq:Bqnum}
  B_{\rm q}(\as,\nf\!=\! 4) \; \cong \; - 0.3183\:\as \:
  ( 1 \: - \: 1.227\: \as \: - \: 3.405\: \as^2 \: + \: \ldots )
  \:\: .
\eeq

\begin{figure}[p]
\vspace{-4mm}
\centerline{\epsfig{file=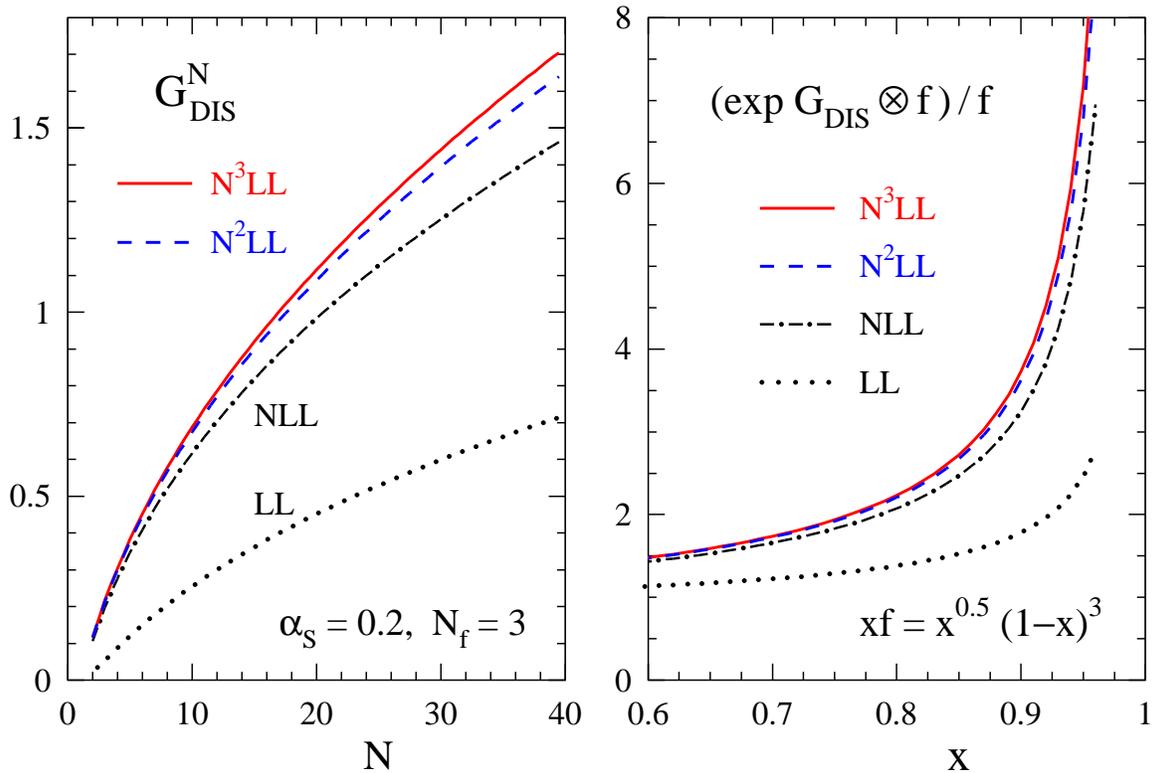,width=15.8cm,angle=0}}
\vspace{-2mm}
\caption{\label{pic:fig1}
  Left: the LL, NLL, N$^2$LL and N$^3$LL approximations for the 
  resummation exponent for standard DIS. Right: the convolutions of
  the exponentiated results with a typical input shape.}
\end{figure}

The ingredients for the resummation of inclusive DIS are now complete, 
and in the left part of Fig.~\ref{pic:fig1} we show the corresponding 
LL, NLL, N$^2$LL and N$^3$LL approximations to the exponent~(\ref
{eq:GNexp}) resulting, for $\as = 0.2$ and three flavours, from 
Eqs.~(\ref{eq:g1n}) -- (\ref{eq:g4n}), (\ref{eq:Aqexp}), 
(\ref{eq:A4pade}) and (\ref{eq:B1}) -- (\ref{eq:B3}). 
For these parameters the expansion (\ref{eq:GNexp}) is stable in the
$N$-range shown in the figure. For example, the relative N$^3$LL
corrections amount to 2\% at $\,N = 10\,$ ($\lambda = 0.33$) and 4\% 
at $\,N = 40\,$ ($\lambda = 0.53$), whereas the corresponding N$^2$LL 
figures read 9\% and 12\%. 
The large third-order contribution to $B_{\rm q}$ actually stabilizes 
$g_4^{}(\lambda)$: for $\,B_{\rm q,3} = 0\,$ the N$^3$LL term at $\,N 
=40\,$ would instead reach 12\%, i.e., the size of the previous order. 
The effect of both $A_{\rm q,4}$ and $\beta_3$, on the other hand, is 
very small, as their respective nullification would change the result 
even at $\,N = 40\,$ by only 0.6\% and 0.1\%.

In the right part of Fig.~1 and in Fig.~2 the exponentiated results
are convoluted with the typical input shape $xf = x^{\,0.5} (1-x)^3$
for a couple of values for $\as$ and $\nf$. The Mellin inversion is in
principle ambiguous due to the Landau poles briefly addressed at the
end of Section 3. We employ the standard `minimal prescription' (thus
adopting the usual fixed-order contour) of Ref.~\cite{Catani:1996xx},
to which the reader is referred for a detailed discussion. For total 
soft-gluon enhancements up to almost an order of magnitude, as shown 
in the figures, the resulting N$^3$LL corrections remain far smaller 
than their N$^2$LL counterparts and amount to less than 10\% even for 
$\as = 0.3$. Note that the dependence on $\nf$ is larger than the 
effect of $g_4^{}$. Thus, at this level of accuracy, a reliable
understanding of heavy-quark mass effects is called for also in the
limit $x \ra 1$. 

\begin{figure}[p]
\vspace{1mm}
\centerline{\epsfig{file=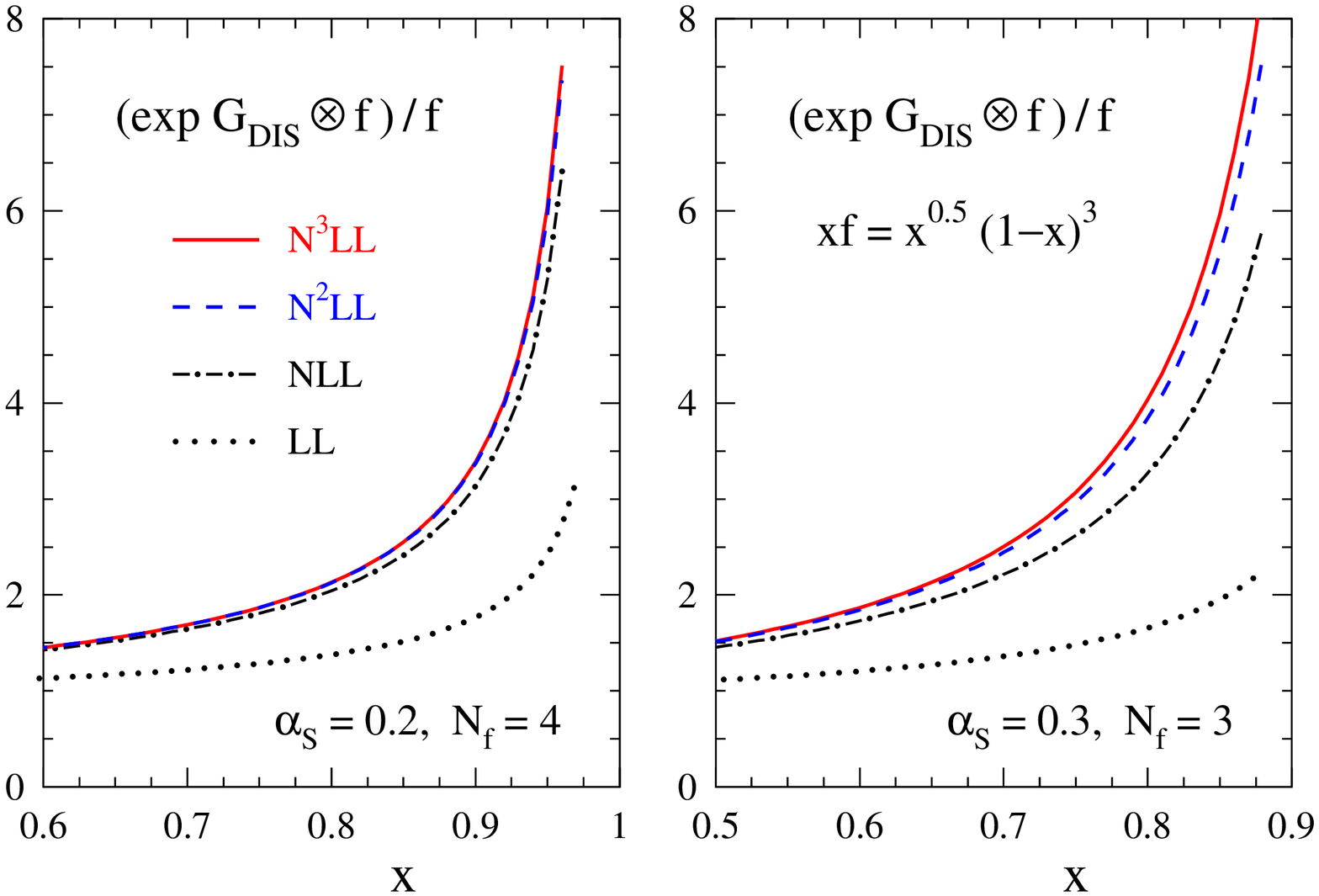,width=15.8cm,angle=0}}
\vspace{-2mm}
\caption{\label{pic:fig2}%
\mbox{As the right part of the previous figure, but for a different 
 value of $\nf$ (left) and $\as$ (right).}} 
\end{figure}

The gluonic coefficients corresponding to Eqs.~(\ref{eq:B1}) --
(\ref{eq:B3}) can be obtained in the same manner from DIS by exchange
of a scalar $\phi$ with a pointlike coupling to gluons, like the Higgs
boson in limit of a heavy top quark. We have derived the corresponding
coefficient functions $c_{\phi,\rm p}^{\,(k)}$ up to $k=3$ already
during the calculations for Ref.~\cite{Vogt:2004mw}, as a process of
this type is required to access the lower row of the flavour-singlet
splitting function matrix. Comparing those results to 
Eqs.~(\ref{eq:cklnN}) for $c_{\phi,\rm g}^{\,(k)}$ yields
\bea
\label{eq:B1g}
 B_{\rm g,1} & = & - \frac{11}{3}\: C_A + \frac{2}{3}\: \nf 
   \; = \; - \beta_0 \:\: ,
   \\[1mm]
\label{eq:B2g}
 B_{\rm g,2} & = &
   C_A^{\,2} \left[ - \frac{611}{9} + \frac{88}{3}\:\z2 
     + 16\:\z3\right] \: + \:
   C_A \nf \left[ \frac{428}{27} - \frac{16}{3}\:\z2 \right] \: + \:
   2\: C_F \nf - \frac{20}{27}\:\nsq \:\: , \quad \\[1mm]
\label{eq:B3g}
 B_{\rm g,3} & = &
   C_A^{\,3} \left[ - \frac{1492081}{1458} + \frac{60875}{81}\:\z2
     + \frac{13796}{27}\:\z3 - \frac{2596}{15}\:\zs 
     - \frac{128}{3}\z2\z3 - 112\:\z5 \right]
     \nn \\[0.5mm] & & \mbox{\hspn} + \:
   C_A^{\,2} \nf \left[ \frac{498329}{1458} - \frac{21014}{81}\:\z2
     - \frac{296}{9}\:\z3 + \frac{568}{15}\:\zs \right] \: - \:
   C_F^{\,2} \nf 
     \nn \\[0.5mm] & & \mbox{\hspn} + \:
   C_A C_F \nf \left[ \frac{8579}{54} - 16\:\z2
     - \frac{832}{9}\:\z3 - \frac{32}{5}\:\zs \right] \: + \:
   C_F \nsq \left[ - \frac{47}{3} + \frac{32}{3}\:\z3 \right] 
     \nn \\[0.5mm] & & \mbox{\hspn} + \:
   C_A \nsq \left[ - \frac{48829}{1458} + \frac{716}{27}\:\z2
     - \frac{176}{27}\:\z3 \right] \: + \:
   \nth \left[ \frac{200}{243} - \frac{8}{9}\:\z2 \right] \:\: ,
\eea
where Eqs.~(\ref{eq:B2g}) and (\ref{eq:B3g}) are new results.
For $\nf = 4$ the numerical expansion of $B_{\rm g}$ reads
\beq
\label{Bgnum}
  B_{\rm g}(\as) \; \cong \; - 0.6631\:\as \:
  ( 1 \: - \: 0.7651\: \as \: - \: 2.696\: \as^2 \: + \: \ldots ) \:\: ,
\eeq
exhibiting an enhanced third order correction similar to that of 
$B_{\rm q}$ in Eq.~(\ref{eq:Bqnum}).

The gluonic threshold resummation resulting from Eqs.~(\ref{eq:Ag}) and
(\ref{eq:B1g}) -- (\ref{eq:B3g}) is illustrated in Fig.~3 using the
practically irrelevant scalar-exchange process, with same parameters as
in Fig.~1 for direct comparison.
The soft and collinear radiation effects are much larger
here due to the larger colour charge of the gluons, but the qualitative
pattern is rather similar to `normal' inclusive DIS.

As mentioned above, the (closely related) Drell-Yan process and Higgs
boson production via gluon-gluon fusion presently represent the only 
other processes for which the NNLL threshold resummation is known, with 
$D_1 = 0$ and \cite{Vogt:2000ci,Catani:2003zt}
\beq
\label{eq:DDYH}
  D_2^{\,\{\rm DY,H\} } \; = \; \{C_F, C_A\} \left[ 
   C_A \left( - \frac{1616}{27} + \frac{176}{3}\,\z2 + 56\,\z3 \right)
   \: + \: \nf \left( \frac{224}{27} - \frac{32}{3}\,\z2 \right) 
   \right] \:\: .
\eeq
Extending a result given in Ref.~\cite{Vogt:2000ci}, we notice the
following conspicuous relation between these coefficients and 
$B_{\rm p,2}\,$:
\bea 
  \frac{1}{2}\: D_2^{\,\rm DY} \: - \: B_{\rm q,2} \: - \: 
  P_{\rm q,\delta}^{\,(1)} &  =  & 7\, \beta_0\, C_F \nn \\
  \frac{1}{2}\: D_2^{\,\rm H}\;\; \: - \: B_{\rm g,2} \: - \: 
  P_{\rm g,\delta}^{\,(1)} &  =  &  \frac{1}{3}\,\beta_0 
     \Big( 4\, C_A + 5\, \beta_0 \Big) \:\: ,
\label{eq:DBPrel}
\eea
where $P_{\rm p,\delta}^{(1)}$ denotes the coefficients of 
$\delta (1-x)$ in the diagonal two-loop splitting functions, and the 
colour structures on the right-hand sides are those of $A_{\rm p,1}$
and $B_{\rm p,1} = - P_{\rm p,\delta}^{(0)}$, multiplied by $\beta_0$.
Note especially the non-trivial cancellation of all $\zeta$-function
terms between the three contributions on the left-hand sides of 
Eqs.~(\ref{eq:DBPrel}) and the vanishing of the right-hand sides for
$\beta_0 \ra 0$.

\begin{figure}[bht]
\vspace{1mm}
\centerline{\epsfig{file=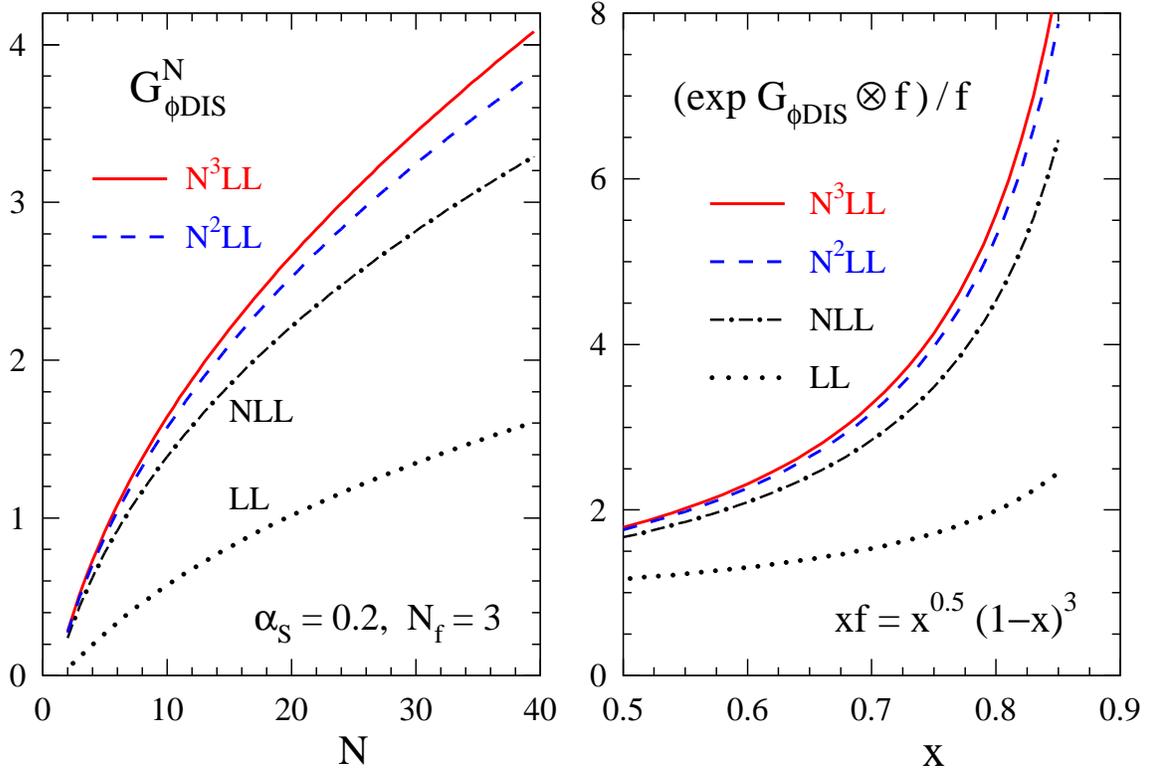,width=15.8cm,angle=0}}
\vspace{-2mm}
\caption{ \label{pic:fig3}
  As Fig.~\ref{pic:fig1}, but for inclusive DIS by exchange of a scalar
  $\phi$ directly coupling to gluons.}
\vspace{9mm}
\end{figure}
%
%
\setcounter{equation}{0}
\section{Fourth-order predictions and tower expansion}
%
%
Another manner to organize the all-order information encoded in Eqs.\
(\ref{eq:cNres}) -- (\ref{eq:Dint}) is to re-expand the exponential,  
\beq
\label{eq:cNexp}
  C^{\,N}(Q^2)/C_{\rm LO}^{\,N}(Q^2) \; =\;  1 + 
  \sum_{k=1}^{\infty} \ars^{\,k} \:
  \sum_{l=1}^{2k} c_{kl}^{} \, \ln^{\,2k-l+1} N \:\: ,
\eeq
and retain only those terms in the second sum which are completely
fixed by the available information on the expansion coefficients in 
Eq.~(\ref{eq:aexp}). Using the notation
\beq
\label{eq:gjexp}
  g_i(\lambda) \; = \; \sum_{k=1}^{\infty} g_{ik} \lambda^k
\eeq
for the expansion of Eqs.~(\ref{eq:g1n}) -- (\ref{eq:g4n}) together
with Eq.~(\ref{eq:aexp}) for $g_0^{}$, the quantities $c_{kl}$ 
in Eq.~(\ref{eq:cNexp}) receive contributions from the following 
coefficients:
\bea
  c_{k1}^{} & : & \quad\:\: g_{11}^{}
  \nn \\[0.5mm]
  c_{k2}^{} & : & \mbox{} +\:\: g_{12}^{} \:\:,\quad g_{21}^{}
  \nn \\[0.5mm]
  c_{k3}^{} & : & \mbox{} +\:\: g_{13}^{} \:\:,\quad g_{22}^{} 
                     \:\:,\quad g_{01}^{}
  \nn \\[0.5mm]
  c_{k4}^{} & : & \mbox{} +\:\: g_{14}^{} \:\:,\quad g_{23}^{} 
                     \:\:,\quad g_{31}^{}
  \nn \\[0.5mm]
  c_{k5}^{} & : & \mbox{} +\:\: g_{15}^{} \:\:,\quad g_{24}^{} 
                     \:\:,\quad g_{32}^{} \:\:,\quad g_{02}^{}
  \nn \\[0.5mm]
  c_{k6}^{} & : & \mbox{} +\:\: g_{16}^{} \:\:,\quad g_{25}^{} 
                     \:\:,\quad g_{33}^{} \:\:,\quad g_{41}^{}
  \nn \\[0.5mm]
  c_{k7}^{} & : & \mbox{} +\:\: g_{17}^{} \:\:,\quad g_{26}^{} 
                     \:\:,\quad g_{34}^{} \:\:,\quad g_{42}^{} 
                     \:\:,\quad g_{03}^{}
  \nn \\[0.5mm]
  c_{k8}^{} & : & \mbox{} +\:\: g_{18}^{} \:\:,\quad g_{27}^{} 
                     \:\:,\quad g_{35}^{} \:\:,\quad g_{43}^{} 
                     \:\:,\quad g_{51}^{} \quad \ldots \quad .
\label{eq:towers}
\eea
The complete relations for the first four terms $c_{k1}^{} \:\ldots\:
c_{k4}^{}$ can be found in Ref.~\cite{Vogt:1999xa} in a slightly
different notation, $g_{31}^{} \ra g_{32}^{}$ (i.e., the second 
index denoting the total power of $\ars$ in Eq.~(\ref{eq:GNexp})).
Note that the  quantities $c_{kl}^{}$ vanish factorially for 
$\,k \ra \infty\,$ and fixed $\,l$. 

Taking into account Eq.~(\ref{eq:DDIS}) and considering the coefficient
$A_i$ as either known or irrelevant, the function $g_i$ for inclusive 
DIS is completely specified by its leading term, obtained by matching 
to the $i$-th order calculation of the coefficient functions as in 
Eqs.~(\ref {eq:cklnN}). The same holds for other processes, at least 
for cases like Eqs.~(\ref {eq:GNdec}), once the required coefficients 
$B_i$ are known from DIS. 

The leading three towers of logarithms, $\,c_{kl}^{}\,$ for any $k$ and 
$l = 1,\:2,\:3\,$, are fixed by a one-loop calculation (providing 
$g_{i1}^{}\,$ for $i = 0,\,1,\,2$) together with the NLL resummation 
(adding $g_{1k}^{}$ and $g_{2k}^{}$ for $k \geq 2\,$). This is the 
status for many important observables 
\cite{Laenen:2004pm,Laenen:2005em}. 
Correspondingly, a two-loop computation of the process under 
consideration specifies $g_{31}^{}$ and $g_{02}^{}$ and hence fixes, 
together with the NNLL resummation, also the next two towers. This is
the accuracy reached for the Drell-Yan process and Higgs production
\cite{Vogt:2000ci,Catani:2003zt}. 
Finally a three-loop computation combined with the N$^3$LL resummation
fixes the first seven towers, $\,c_{kl}^{}\,$ for $l = 1, \:\ldots, 
\:7\,$. With the results of Ref.~\cite{Vermaseren:2005qc} and Sections
3 and 4, we have now reached this point for the structure function 
$F_2$ in DIS.

The resulting four-loop predictions, in $x$-space expressed in terms of 
the coefficients of the +-distributions $\DD_k = [ (1-x)^{-1} \ln (1-x)
]_+$, for the six highest terms read   
\bea
\label{eq:cq4d7}
  c_{\,2,\rm q}^{\,(4)} \Big|_{\,\DD^{}_7}  &\!\! = \! &
    \frac{16}{3}\: C_F^{\,4} \:\: , \\[2mm]
\label{eq:cq4d6}
  c_{\,2,\rm q}^{\,(4)} \Big|_{\,\DD^{}_6}  &\!\! = \! &
  - 28\: C_F^{\,4} - \frac{308}{9}\: C_A C_F^{\,3}
  + \frac{56}{9}\: C_F^{\,3}\nf  \:\: , \\[2mm]
\label{eq:cqd5}
  c_{\,2,\rm q}^{\,(4)} \Big|_{\,\DD^{}_5}  &\!\! = \! &
  C_F^{\,4} \Big[ - 18 - 128\:\z2 \Big] \: + \: 
  C_A C_F^{\,3} \left[ \frac{998}{3} - 48\:\z2 \right] \: + \:
  \frac{1936}{27}\: C_A^{\,2} C_F^{\,2} 
  \nn \\[0.5mm] & & \mbox{\hspn} - \:
  \frac{164}{3}\: C_F^{\,3} \nf \: - \: 
  \frac{704}{27}\: C_A C_F^{\,2} \nf \: + \:
  \frac{64}{27}\: C_F^{\,2} \nsq \:\: , \\[2mm]
\label{eq:cqd4}
  c_{\,2,\rm q}^{\,(4)} \Big|_{\,\DD^{}_4} &\!\! = \! &
  C_F^{\,4} \,\left[ 210 + 600\:\z2 + \frac{400}{3}\:\z3 \right] \: + \:
  C_A  C_F^{\,3} \,\left[ -\frac{27835}{27} + \frac{6800}{9}\z2
    + 400\:\z3 \right] \nn \\[0.5mm] & & \mbox{\hspn} + \:
  C_A^{\,2} C_F^{\,2} \,\left[ -\frac{24040}{27} + \frac{440}{3}\:\z2 
    \right] \: - \: 
  \frac{1331}{27}\: C_A^{\,3} C_F \: + \: 
  C_F^{\,3} \nf \,\left[ \frac{4630}{27} - \frac{1040}{9}\:\z2 \right]
     \nn \\[0.5mm] & & \mbox{\hspn} + \:
  C_A C_F^{\,2} \nf \,\left[ \frac{8120}{27} - \frac{80}{3}\:\z2 \right]
     \: + \: 
  \frac{242}{9}\: C_A^{\,2} C_F \nf \: - \: 
  \frac{640}{27}\: C_F^{\,2} \nsq \: - \: 
  \frac{44}{9}\: C_A C_F \nsq 
     \nn \\[0.5mm] & & \mbox{\hspn} + \:
  \frac{8}{27}\: C_F n^{\,3}_{\! f} \:\: , \\[2mm]
\label{eq:cqd3}
  c_{\,2,\rm q}^{\,(4)} \Big|_{\,\DD^{}_3} &\!\! = \! &
  C_F^{\,4} \,\left[ \frac{113}{2} + 264\:\z2 - 1072\:\z3
      + \frac{1392}{5}\:\zs \right] \: + \: 
  C_A^{\,3} C_F \,\left[ \frac{55627}{81} - \frac{968}{9}\:\z2 \right] 
     \nn \\[0.5mm] & & \mbox{\hspn} + \:
  C_A C_F^{\,3} \,\left[ - \frac{1534}{3} - \frac{41824}{9}\:\z2 
      - \frac{8800}{9}\:\z3 + \frac{3128}{5}\:\zs \right] 
      \nn  \\[0.5mm] & & \mbox{\hspn} + \:
  C_A^{\,2} C_F^{\,2} \,\left[ \frac{2154563}{486} 
      - \frac{52912}{27}\:\z2 - \frac{13024}{9}\:\z3 
      + \frac{864}{5}\:\zs \right] 
     \nn \\[0.5mm] & & \mbox{\hspn} + \:
  C_F^{\,3} \nf\, \left[ - \frac{280}{3} + \frac{7216}{9}\:\z2 
      + \frac{1888}{9}\:\z3 \right] \: + \: 
  C_A^{\,2} C_F \nf\, \left[ - \frac{9502}{27} + \frac{352}{9}\:\z2 
      \right] \nn \\[0.5mm] & & \mbox{\hspn} - \:
  C_A C_F^{\,2} \nf\, \left[ \frac{339134}{243} 
      - \frac{14096}{27}\:\z2 - \frac{1216}{9}\:\z3 \right] \: + \:
  C_A C_F \nsq \left[ \frac{1540}{27} - \frac{32}{9}\:\z2 \right]
      \nn \\[0.5mm] & & \mbox{\hspn} + \:
  C_F^2 \nsq \left[ \frac{24238}{243} - \frac{928}{27}\:\z2 \right]
     \: - \:
  \frac{232}{81}\: C_F n^{\,3}_{\! f} 
      \:\: , \\[3mm]
  c_{\,2,\rm q}^{\,(4)} \Big|_{\,\DD^{}_2} &\!\! = \! &
  C_F^{\,4} \,\left[ - \frac{1299}{2} - 2808\:\z2 + 1392\:\z3 
      - 1836\:\zs - 640\:\z2\z3 + 4128\:\z5 \right]  
      \nn \\[0.5mm] & & \mbox{\hspn} + \:
  C_A C_F^{\,3} \,\left[ \frac{13990}{3} + \frac{30704}{3}\:\z2
      + \frac{2716}{3}\:\z3 - \frac{12906}{5}\:\zs - 3648\:\z2\z3 
      - 720\:\z5
      \right] \nn \\[0.5mm] & & \mbox{\hspn} - \:
  C_A^{\,2} C_F^{\,2} \,\left[ \frac{2254339}{243} 
      - \frac{86804}{9}\:\z2 - \frac{24544}{3}\:\z3 
      + \frac{4034}{3}\:\zs + 832\:\z2\z3 + 1392\:\z5 \right]  
      \nn \\[0.5mm] & & \mbox{\hspn} + \:
  C_A^{\,3} C_F \,\left[ - \frac{649589}{162} + \frac{4012}{3}\:\z2
      + 1452\:\z3 - \frac{968}{5}\:\zs \right]  
      \nn \\[0.5mm] & & \mbox{\hspn} + \:
  C_A C_F^{\,2} \nf\, \left[ \frac{713162}{243} - \frac{82004}{27}\:\z2
      - \frac{10600}{9}\:\z3 + \frac{3772}{15}\:\zs\right] 
      \nn \\[0.5mm] & & \mbox{\hspn} + \:
  C_A^{\,2} C_F \nf\, \left[ \frac{17189}{9} - \frac{5096}{9}\:\z2
      - 352\:\z3 + \frac{176}{5}\:\zs\right] 
      \\[0.5mm] & & \mbox{\hspn} + \:
  C_F^{\,3} \nf\, \left[ - \frac{145}{9} - \frac{5132}{3}\:\z2
      - 936\:\z3 + \frac{1032}{5}\:\zs\right] \: + \: 
  C_F n^{\,3}_{\! f}  \left[ \frac{940}{81} - \frac{32}{9}\:\z2
      \right] \nn \\[0.5mm] & & \mbox{\hspn} - \:
  C_A C_F \nsq \left[ \frac{7403}{27} - \frac{688}{9}\:\z2
      - 16 \:\z3 \right] \: - \:
  C_F^{\,2} \nsq \left[ \frac{52678}{243} - \frac{6104}{27}\:\z2
      - \frac{304}{9}\:\z3 \right] \:\: . \nn
\eea
The seventh term with $\DD_{1}$ is not exactly known, since the 
fourth-order contribution to $A_{\rm q}$ has not been computed so far. 
Inserting the numerical values for the $\zeta$-functions and the QCD 
colour factors, including $d^{abc}d_{abc}/n_c = 5\nf / 18$, the 
resummation prediction is given by
\beq
  c_{\,2,\rm q}^{\,(4)} \Big|_{\,\DD^{}_1}  = \;
  - 286702 + 64219.0\:\nf - 2019.24\:\nsq + 2.0166\:n^{\,3}_{\! f}
  - 63.402 \:\fl11\,\nf + A_{\rm q,4} \:\: .  
\eeq
As mentioned above, numerically insignificant are both the uncertainty
due to $A_{\rm q,4}$ (estimated in Eq.~(\ref{eq:A4pade})) and the
singlet$\,/\,$non-singlet difference introduced by the $\fl11$ 
contribution of Eq.~(\ref{eq:g03dis}).
It is also interesting to note that the fourth coefficient $\beta_3$ of 
the beta function~\cite{vanRitbergen:1997va,Czakon:2004bu} with its
quartic group invariants $d_F^{\,abcd}$ and $d_A^{\,abcd}$ only enters 
the eighth tower, starting at the fifth order in $\as$. 

\begin{table}[htb]
\begin{center}
\begin{tabular}{|r||r|r|r|r|r|r|r|}\hline
   &        &        &        &        &            \\[-4mm]
$k$& $c_{k1}^{} \quad $ & $c_{k2}^{} \quad $ &
     $c_{k3}^{} \quad $ & $c_{k4}^{} \quad $ &
     $c_{k5}^{} \quad $ & $c_{k6}^{} \quad $ &  
     $c_{k7}^{} \quad $ \\[1mm]
                                                     \hline \hline
   &         &         &         &         &         & & \\[-4mm]
 1 & 2.66667 &  7.0784 &   ---   &   ---   &   ---   &   ---  &   ---   \\[1mm]
 2 & 3.55556 & 26.2834 &  40.760 &$-$67.13 &   ---   &   ---  &   ---   \\[1mm] 
 3 & 3.16049 & 44.9210 & 238.885 &  470.82 &$-$620.3 &$-$1639 &   ---   \\[1mm]
 4 & 2.10700 & 48.7090 & 477.854 & 2429.46 &  5240.0 &$-$1824 &$-$30318 \\[1mm]
 5 & 1.12373 & 38.3254 & 581.518 & 5015.18 & 25150.5 &  48482 &   11268 \\[1mm]
 6 & 0.49944 & 23.5617 & 505.972 & 6432.95 & 52129.7 & 225320 &  675012 \\[1mm]
 7 & 0.19026 & 11.8592 & 340.954 & 5933.61 & 68602.9 & 485712 & 2494841 \\[1mm]
 8 & 0.06342 &  5.0463 & 186.822 & 4244.45 & 65550.0 & 668223 & 4979993 \\[1mm]
 9 & 0.01879 &  1.8583 &  86.041 & 2467.72 & 48805.8 & 666670 & 6718531 \\[1mm]
10 & 0.00501 &  0.6028 &  34.118 & 1204.34 & 29604.7 & 517490 & 6747332 \\[1mm]
                                                     \hline
\end{tabular}
\vspace{3mm}
\caption{Numerical values of the four-flavour coefficients $c_{kl}^{}$ 
 in Eq.~(\ref{eq:cNexp}) for the quark coefficient function in DIS. The 
 first six columns are exact up to the numerical truncation, and the 
 same for $F_1$, $F_2$ and $F_3$. The seventh column refers to $F_1$ and
 $F_2$ for the photon-exchange flavour-singlet case, $\fl11 = 1/10$
 \cite{vanRitbergen:1997va}, and uses the estimate (\ref{eq:A4pade}) for
 the four-loop cusp anomalous dimension.} 
\vspace*{-4mm}
\end{center}
\end{table}
 
The numerical values of the $N$-space coefficients $c_{kl}^{}$ in 
Eq.~(\ref{eq:cNexp}) are presented in Table~1 for $l \leq 7$ and $k\leq 
10$. Recall that also these coefficients refer to an expansion in $\ars
= \as/(4\pi)$. Whatever the normalization of the expansion parameter, 
however, the coefficients in each column (tower) finally vanish for $k 
\ra \infty$, as mentioned below Eq.~(\ref{eq:towers}). Thus the series 
(\ref{eq:cNexp}) converges at all $N \neq 0$ for any finite number of
towers $\,l_{\rm max}$, i.e., with the upper limit in the second sum
replaced by $\,l_{\rm max}$. The Mellin inversion of the product with
the parton distributions $f^{N}$ is therefore well-defined, in contrast 
to the fully exponentiated result discussed above.

Before we turn to the higher-order predictions, it is instructive to
compare the approximations by the leading large-$x$ and large-$N$ terms 
to the completely known two- and three-loop coefficient functions
\cite{vanNeerven:1991nn,Moch:1999eb,Vermaseren:2005qc}. 
This is done in Fig.~4 for the successive approximations in terms of 
the \mbox{+-distributions} $\DD_k$ defined above Eq.~(\ref{eq:cq4d7}). 
The corresponding results for the expansion in powers of $\,\ln\, N\,$ 
are presented in Fig.~5.
Obviously both expansions reproduce the exact large-$x$ behaviour (up 
to terms not increasing as $x \!\ra\! 1$ for the ratios shown in the 
figures) at order $\as^{\,n}$ once all enhanced terms, $\DD_k$ with 
$k = 0,\,\ldots,\, 2n\!-\!1$ or $\ln^{\,l}N$ with $l = 1,\,\ldots,\,2n$,
have been taken into account. 
The \mbox{$x$-space} expansion, however, would lead to a gross 
overestimate if only the terms up to $k\simeq n\!+\!1$ were known. 
The convergence in the $N$-space approach, on the other hand, is much 
smoother, with a good approximation already reached after $n$ terms.
 
\begin{figure}[p]
\vspace{-4mm}
\centerline{\epsfig{file=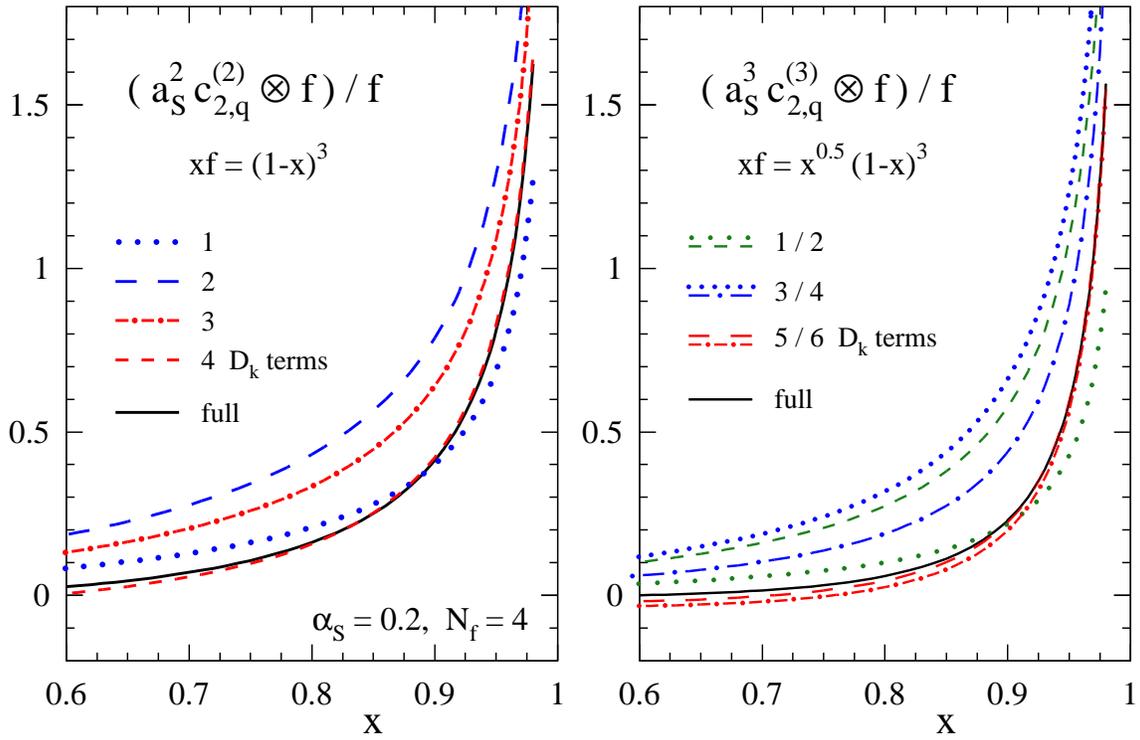,width=15.6cm,angle=0}}
\vspace{-3mm}
\caption{ \label{pic:fig4}
 Convolution of the two-loop (left) and three-loop (right) contributions
 to the DIS coefficient functions $C_{2,\rm q}$ with a typical input 
 shape. Shown are the full results 
 \cite{vanNeerven:1991nn,Moch:1999eb,Vermaseren:2005qc} and the
  large-$x$ expansion by successively including the +-distributions 
  $\DD_k$, respectively starting with $\DD_3$ and $\DD_5$.} 
\end{figure}
\begin{figure}[p]
\vspace{2mm}
\centerline{\epsfig{file=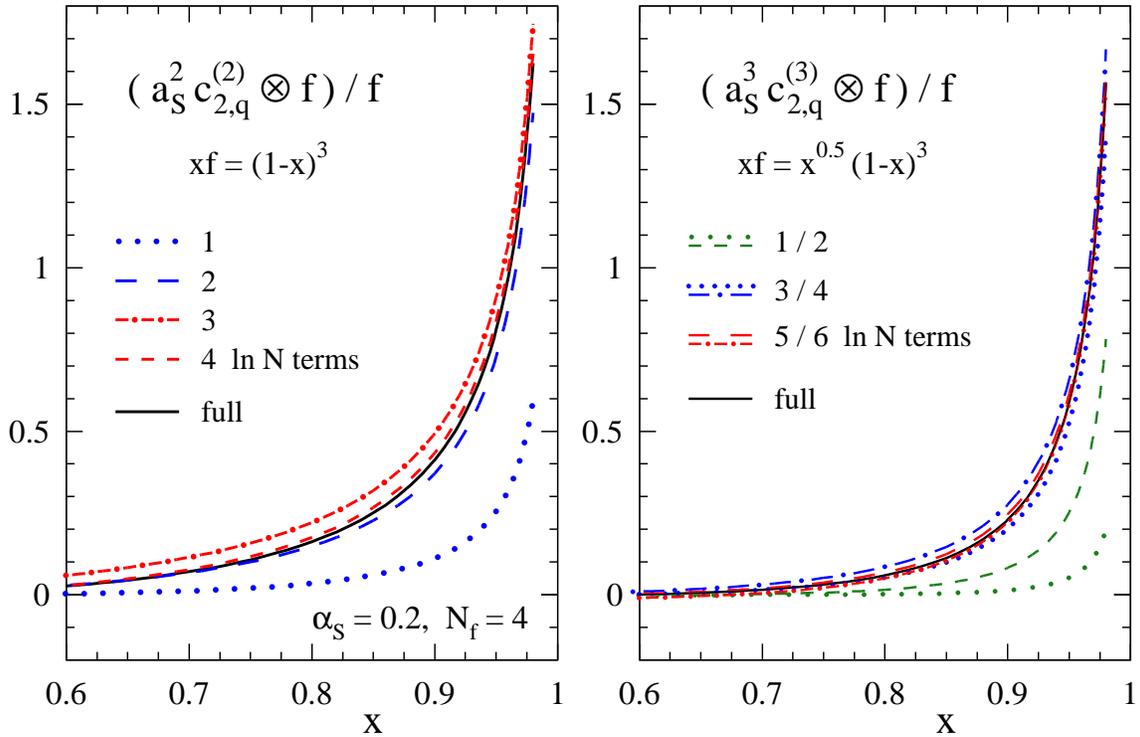,width=15.6cm,angle=0}}
\vspace{-3mm}
\caption{ \label{pic:fig5}
 As the previous figure, but for large-$N$ expansion in terms of 
 decreasing powers
 of $\ln\, N\,$.}
\end{figure}

A similar pattern is found for the fourth-order coefficient function
$c_{\,2,\rm q}^{\,(4)\,}$ illustrated in the same manner in the left
part of Fig.~6: the expansion in decreasing powers of $\ln\, N$
stabilizes after the fourth term. Based on these results and the
higher-order coefficients shown in Table~1, we expect that the first
$l$ logarithms should provide a good estimate up to about the $l$-th 
order in $\as$, but severely underestimate the effect of the 
coefficient functions of much higher orders. 
Consequently, the tower expansion should underestimate the corrections 
towards $x \!\ra\! 1$, where more and more orders become relevant. This 
is exactly the pattern shown in the right part of Fig.~6, where the 
predictions of all effects beyond order $\as^{\,3}$ are compared between
the tower expansion and the full exponentiation (for the latter again
using a `minimal-prescription' contour~\cite{Catani:1996xx}). Both 
approaches agree very well, for the chosen input parameters, at 
$x < 0.93$, but start to diverge at $x \gsim 0.95$ where the 
exponentiation is also intrinsically more stable.

\begin{figure}[htb]
\vspace{2mm}
\centerline{\epsfig{file=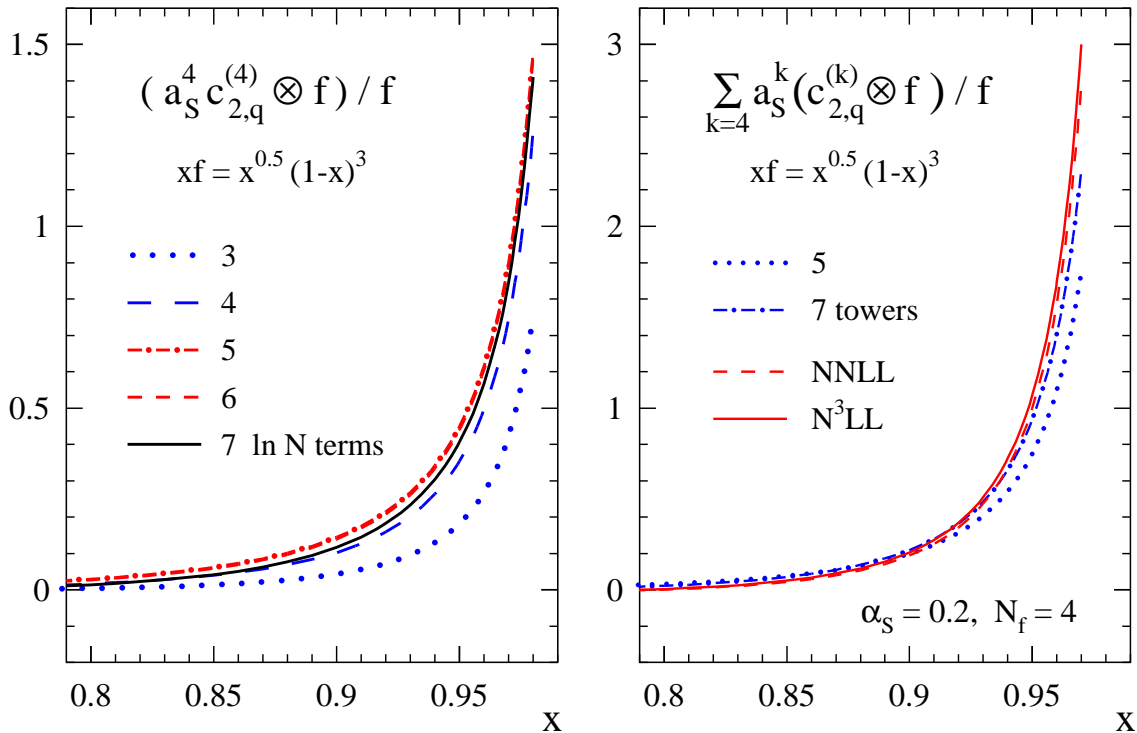,width=15.6cm,angle=0}}
\vspace{-3mm}
\caption{ \label{pic:fig6}
 Left: the successive approximations of the four-loop coefficient 
 function $c_{\,2,\rm q}^{\,(4)\,}$ by the \mbox{large-$N$} terms 
 specified in Table~1, illustrated by the convolution with a typical 
 quark distribution. 
 Right: corresponding results for the effect of all orders beyond 
 $\as^{\,3}$ as obtained from the tower expansion with up to seven 
 towers and from the exponentiation up to N$^3$LL accuracy.}
\vspace{2mm}
\end{figure}
%
%
\section{Summary}
%
%
We have extended the threshold resummation exponents
\cite{Sterman:1987aj,Catani:1989ne,Vogt:2000ci,Catani:2003zt} 
for few-parton processes to the fourth logarithmic (N$^3$LL) order
collecting the terms $\as^{\,2} (\as \ln N)^k$ to all orders in $\as$.
For our reference process, inclusive deep-inelastic scattering (DIS),
the N$^3$LL contributions are specified by two universal expansion
parameters: the four-loop cusp anomalous dimension $A_{\rm q,4}$ and
the third-order quantity $B_{\rm q,3}$ which defines the jet function 
resumming collinear radiation off an unobserved final-state quark.
The former coefficient has not been computed so far, but can be safely
expected to have a very small effect of less than 1\%. In fact, the 
perturbative expansion up to $A_3$ \cite{Moch:2004pa,Vogt:2004mw} 
does not exhibit enhanced higher-order corrections, and $A_4$ can be 
estimated by Pad\'e approximations. We have calculated the more
important second coefficient $B_{\rm q,3}$ by comparison of the 
expanded resummation result to our recent third-order calculation of
electromagnetic DIS~\cite{Vermaseren:2005qc}. 

The perturbative expansion of $B_{\rm q}$ seems to indicate, as far as
this can be judged from the first three terms, the onset of a factorial
enhancement of the higher-order coefficients. However, the rather large 
size of the coefficient $B_{\rm q,3}$ actually stabilizes the 
logarithmic expansion of the coefficient functions. In fact, the 
N$^3$LL corrections are very small at large scales $Q^2$, and even 
facilitate a reliable prediction of the soft-gluon effects at scales as
low as $Q^2 \approx 4 \mbox{ GeV}^2$ (corresponding to $\as \:\simeq\: 
0.3)$ down to very small invariant masses $W$ of the hadronic final 
state in $ep \ra eX$, $\,W^2 - m_p^2 \:\approx\: 0.5 \mbox{ GeV}^2$. 
Thus we expect our results to be useful also for low-scale data 
analyses using parton-hadron duality concepts.

The threshold resummation can also be employed to predict, order by 
order in $\as$, the leading $\,\ln\,N\,$ contributions to the 
higher-order coefficient functions. At the level of accuracy reached in 
the present article for inclusive DIS, the exponentiation fixes the
seven highest terms, $\ln^n N$ with $2l\!-\!6 \leq n \leq 2k\,$, at all
orders $k \geq 4$ of $\as$. Already the highest $k$ powers of $\ln\, N$
provide a good estimate of the soft-gluon enhancement of the $k$-loop 
coefficient functions at least for $k \leq 7$, in contrast to the 
(expected, see Ref.~\cite{Catani:1996xx}) worse behaviour of the 
corresponding expansion in $x$-space +-distributions. Except very close 
to threshold, where too many orders in $\as$ become important, the 
summation of the above seven $N$-space logarithms to all orders yields 
a good agreement with the exponentiated coefficient function. This 
agreement further confirms the `minimal prescription'
\cite{Catani:1996xx} used for defining the in principle ambiguous 
Mellin inversion of the resummation exponential.

Besides the standard (gauge-boson exchange) process, we have also 
considered DIS by exchange of a scalar directly coupling to gluons.
By comparison of the resummation to our unpublished three-loop 
coefficient function for this process we have derived, for the first
time, the second and third order contributions to the coefficient 
$B_{\rm g}$ governing the jet function of a final-state gluon.
The quantity $B_{\rm g,2}$ will be employed to extend the NNLL 
resummation to more processes, once the corresponding NNLO results
required to fix the process-dependent large-angle soft contribution
become available. Finally we would like to draw attention to the 
curious relation (\ref{eq:DBPrel}) which connects, for both the quark 
and gluon channels, the second-order splitting functions, jet functions 
and the two-parton (Drell-Yan) large-angle soft emissions in a 
non-trivial manner.
%
%
\subsection*{Acknowlegments}
%
%
We thank E.~Laenen and W.~Vogelsang for stimulating discussions. 
The work of S.M. has been supported in part by the Helmholtz 
Gemeinschaft under contract VH-NG-105 and by the Deutsche 
Forschungsgemeinschaft in Sonderforschungsbereich/Transregio 9.
The work of J.V. has been part of the research program of the
Dutch Foundation for Fundamental Research of Matter (FOM).
%
%
\renewcommand{\theequation}{A.\arabic{equation}}
\setcounter{equation}{0}
\section*{Appendix}
%
%
Here we show some key elements of the calculation of the resummation
exponents $g_i$ presented in Section 3. Useful auxiliary relations 
(for $|\,x\,| < 1$) are 
\bea
\label{eq:1mxexp}
{1 \over (1-x)^{n-\epsilon}} &=& 
\sum\limits_{i=0}^\infty\, {\Gamma(n-\epsilon+i) \over 
\Gamma(n-\epsilon)}\, {x^i \over i!}  \:\: ,
\\
{\ln^k(1-x) \over (1-x)^{n-\epsilon}} & = &
\left( {\partial \over \partial \epsilon} \right)^{\! k} 
{1 \over (1-x)^{n-\epsilon}} \:\: .
\eea
The Mellin transforms of the +-distributions follow from the results 
for harmonic polylogarithms \cite{Remiddi:1999ew} and are given by
\beq
\label{eq:auxMtrf}
\int\limits_0^1\, dz\, {z^{\,N}-1 \over 1-z}\, \ln^k(1-z) 
\; = \; (-1)^{k+1}\, k!\, S_{\underbrace{1,\dots,1}_{k+1}}(N) \:\: ,
\eeq
where $S_{m_1,\dots,m_k}(N)$ denotes the harmonic sums 
\cite{Vermaseren:1998uu}.  Eqs.~(\ref{eq:1mxexp}) -- (\ref{eq:auxMtrf})
lead to the master formula for the derivation of the functions $g_i$,
\bea
\lefteqn{
\int\limits_0^1\, dz\, {z^{\,N}-1 \over 1-z}\, 
{1 \over (1 + a \ln(1-z))^{n-\epsilon}} \; = } 
\nonumber \\
&\displaystyle
 - \sum\limits_{i=0}^\infty\, {\Gamma(n-\epsilon+i) \over 
\Gamma(n-\epsilon)} 
\, (a_s\beta_0)^i \,S_{\underbrace{1,\dots,1}_{i+1}}(N)\,
\sum\limits_{j=0}^\infty\, \left(
\begin{array}{c} \! i+j-1 \! \\ j \end{array}
\right) \left(-a_s\beta_0\ln{Q^2 \over \mu_r^2}\right)^j
\label{eq:gimaster}
\eea
with $a=(a_s\beta_0)/(1+a_s\beta_0\ln Q^2/\mu_r^2)$.
The double sum in Eq.~(\ref{eq:gimaster}) can be solved to the desired 
logarithmic accuracy with the algorithms for the summation of nested 
sums~\cite{Moch:2001zr} coded in {\sc Form}~\cite{Vermaseren:2000nd}. 
The expansion of the Gamma function in powers of $\epsilon$ for positive
integers $n$ reads
\bea
\label{eq:expgamma}
{ \Gamma(n+1+\epsilon) \over  n!\, \Gamma(1+\epsilon)} &=& 
         1
       + \epsilon \* 
            S_{1}(n)
       + \epsilon^2 \* \bigl(
            S_{1,1}(n)
          - S_{2}(n)
          \bigr)
       + \epsilon^3 \* \bigl(
            S_{1,1,1}(n)
          - S_{1,2}(n)
\nonumber\\[-1mm]
& &\mbox{}
          - S_{2,1}(n)
          + S_{3}(n)
          \bigr)
       + \epsilon^4 \* \bigl(
            S_{1,1,1,1}(n)
          - S_{1,1,2}(n)
          - S_{1,2,1}(n)
\nonumber\\[1mm]
& &\mbox{}
          + S_{1,3}(n)
          - S_{2,1,1}(n)
          + S_{2,2}(n)
          + S_{3,1}(n)
          - S_{4}(n)
          \bigr)
+ {\cal O}(\epsilon^5) \:\: .
\eea
Finally, with $\theta_{ij} = 1$ for $i\geq j$ and $\theta_{ij} = 0$ 
else, the sums $S_{1,\dots,1}(N)$ are factorized according to
\bea
i! \: S_{\underbrace{1,\dots,1}_{i}}(N) & \! = \! & 
(S_1(N))^{i} 
+ {1 \over 2} i(i-1) S_2(N) (S_1(N))^{i-2}
+ {1 \over 3} i(i-1)(i-2) S_3(N) (S_1(N))^{i-3} 
\nonumber\\[-3mm]
& & \mbox{} 
+ {1 \over 4} i(i-1)(i-2)(i-3) \left(S_4(N)
+ {1 \over 2} (S_2(N))^2\right) (S_1(N))^{i-4}
+ \dots 
\\[1mm]
&\!\simeq\!& 
\theta_{i1} \ln^{i} \tilde{N}
+ {1 \over 2} \theta_{i3} i(i-1) \z2 \ln^{i-2} \tilde{N}
+ {1 \over 3} \theta_{i4} i(i-1)(i-2) \z3 \ln^{i-3} \tilde{N}
\nonumber\\[1mm]
& &
+ {1 \over 4} \theta_{i5} i(i-1)(i-2)(i-3) \left(\z4+{1 \over 2} 
\z2^2\right) \ln^{i-4} \tilde{N}
+ \dots \:\: ,
\eea
where $\tilde{N} = N e^{\,\gamma_e}$ and the algebraic properties of harmonic 
sums have been used~\cite{Vermaseren:1998uu}.

\end{document}